\documentclass{article}

\usepackage{PRIMEarxiv}

\usepackage[utf8]{inputenc} 
\usepackage[T1]{fontenc}    
\usepackage{hyperref}       
\usepackage{url}            
\usepackage{booktabs}       
\usepackage{amsfonts}       
\usepackage{nicefrac}       
\usepackage{microtype}      
\usepackage{lipsum}
\usepackage{fancyhdr}       
\usepackage{graphicx}       
\graphicspath{{media/}}     
\usepackage{multirow}
\usepackage{enumitem}
\usepackage[acronym]{glossaries}
\usepackage[dvipsnames]{xcolor}
\usepackage[numbers]{natbib}
\usepackage{adjustbox}
\usepackage{multirow}
\usepackage[normalem]{ulem}
\useunder{\uline}{\ul}{}

\usepackage{amsmath,amssymb,amsfonts}
\usepackage{algorithmic}
\usepackage{graphicx}
\usepackage{textcomp}
\usepackage{xcolor}
\usepackage{svg}
\usepackage{float}
\usepackage{stfloats}
\usepackage{natbib}
\usepackage{nomencl}
\makenomenclature
\usepackage[acronym]{glossaries}
\usepackage{url}

\usepackage{orcidlink}
\usepackage{enumitem}

\usepackage{booktabs}
\usepackage{graphicx}  
\usepackage{multirow}
\newcommand{\yes}{\checkmark}

\makeglossaries
\newacronym{vd}{VD}{Voronoi Diagram}
\newacronym{gpm}{GPM}{Grid Point Modeling}
\newacronym{gnn}{GNN}{Graph Neural Networks}
\newacronym{ntl}{NTL}{Nighttime Light}
\newacronym{hgnn}{HGNN}{Heterogeneous Graph Neural Network}
\newacronym{gdp}{GDP}{Gross Domestic Product}
\newacronym{gva}{GVA}{Gross Value Added}
\newacronym{itl}{ITL}{International Territorial Level}
\newacronym{rci}{RCI}{Residential + Commercial + Industrial}
\newacronym{hgt}{HGT}{Heterogeneous Graph Transformer}
\newacronym{rmse}{RMSE}{Root Mean Square Error}
\newacronym{mae}{MAE}{Mean Absolute Error}
\newacronym{corr}{Corr}{Pearson Correlation Coefficient}

\title{Mechanism-Dependent Antagonism of Auxiliary Information in Substation-Level Load Disaggregation for Distribution Network Planning
}

\author{
  Xuanhao Mu, Kundan Thota, Nan Liu, Jianlei Liu, Thorsten Schlachter and Veit Hagenmeyer \\
  Institute for Automation and Applied Informatics (IAI) \\
  Karlsruhe Institute of Technology \\
  76344 Eggenstein-Leopoldshafen, Germany\\
  \texttt{xuanhao.mu@kit.edu kunda.thota@kit.edu nan.liu@student.kit.edu}\\
  \texttt{jianlei.liu@kit.edu thorsten.schlachter@kit.edu veit.hagenmeyer@kit.edu} \\
}

\begin{document}
\maketitle

\begin{abstract}
Open-source energy system models disaggregate zonal electricity demand to substations through Voronoi-based preprocessing pipelines that combine socioeconomic weighting with auxiliary spatial corrections.
Whether the same auxiliary data helps or harms when the weighting component shifts from rule-based to learned has not been investigated.
We fix Voronoi partitioning and cross two design axes on metered demand from 1{,}891 British primary substations: the demand-weighting method (static Grid Point Model versus a heterogeneous graph neural network trained under self-supervised land-use reconstruction) and the mechanism through which Nighttime Light (NTL) intensity and substation-proximity signals enter the allocation (multiplicative post-correction applied after inference versus prior-loss terms injected during training), giving 15 configurations.
Mechanism-isolation experiments further test additive post-correction and random-noise controls to pinpoint the structural cause of any performance reversal.
The same auxiliary data reduces RMSE by $41\,\%$ on the static base but increases it by $21\,\%$ on the GNN base under multiplicative post-correction ($p < 0.001$ for both); the best static pipeline outperforms the best GNN variant by $19\,\%$.
Post-correction on the GNN improves rank-order correlation ($p < 0.001$) yet worsens absolute error, so correlation-only evaluation masks the calibration penalty.
The isolation experiments trace this reversal to the multiplicative correction form under demand conservation constraints, not to signal redundancy; switching to additive post-correction eliminates the antagonism entirely.
A transfer check on 13 German primary substations confirms directional replication and shows amplified antagonism where the GNN baseline already explains over $95\,\%$ of demand variance.
The NTL and proximity signals behind the $41\,\%$ static improvement are publicly available at no cost and should be adopted as default corrections in static pipelines; method evaluation should report RMSE and correlation jointly, as the two metrics diverge under post-correction on learned representations.
\end{abstract}


\section{Introduction}
\label{sec:introduction}

Distribution network planning requires substation-level demand estimates, yet electricity consumption data is typically available only at the zonal or regional level. Open-source energy system models address this mismatch through a Voronoi-based spatial allocation pipeline~\cite{Horsch2018pypsa, Hulk2017allocation}.
This upstream module consists of three configurable steps---Voronoi partitioning, socioeconomic demand weighting, and auxiliary spatial corrections---and is implemented in both open-source frameworks and commercial planning tools~\cite{Raventos2022comparison}.
Multiplicative post-correction, in which agent-level demands are scaled by auxiliary factors and renormalized to conserve regional totals, is the de facto standard for the third step.
Its design presupposes that the base allocation carries little spatial information, which is a condition met by uniform or proxy-rule weighting but not examined for learned alternatives.

Within this pipeline, the partitioning is typically fixed by a \gls{vd}~\cite{crato_figuring_2010}.
Weighting is handled either by static rules or neural networks.
Auxiliary information such as \gls{ntl} intensity~\cite{Chen2022global} and substation proximity (Proximity)~\cite{Borges2020enhancing}, is attached last as a multiplicative correction layer.
Since each step is configured independently, the pipeline implicitly assumes plug-compatibility: a better weighting component can be substituted while the same auxiliary correction is reused unchanged.

Although prior reviews~\cite{Raventos2022comparison, Patil2024systematic} and single-axis comparisons~\cite{Hulk2017allocation, Mu2026Improving, Singh2015high} document wide variation in weighting and correction practices, none of these studies test whether the correction mechanism remains appropriate when the weighting component is replaced.
Moreover, they do not distinguish whether the same signal enters as a post-correction or as a training-time prior. 
Therefore, they cannot determine whether changing the integration mechanism preserves or reverses model ranking.
The question is not specific to energy: identical domain knowledge produces different model behaviors when encoded as soft constraints, architectural constraints, or auxiliary losses~\cite{Chen2024Physics-Informed, Liang2022Knowledge, Gong2019acomparison}, yet spatial disaggregation has not been tested for this mechanism sensitivity under multiplicative correction with demand conservation.

The present paper asks whether the same auxiliary spatial information helps or harms once the weighting component changes from rule-based to learned.
The main contributions are:
\begin{enumerate}[label=(\arabic*),leftmargin=*]
\item \textbf{Mechanism-dependent failure mode identified and attributed.}

The same \gls{ntl} and Proximity data reduces \gls{rmse} by $41\%$ on a static base but increases it by $21\%$ on a \gls{gnn} base under multiplicative post-correction ($p < 0.001$ for both), reversing model ranking so that the best static pipeline outperforms the best \gls{gnn} variant in \gls{rmse} by $19\%$ (GPMpostNP vs.\ GNNpriorNP, $(9.03-7.31)/9.03$).
Mechanism-isolation experiments attribute the reversal to the structural interaction between multiplicative correction and demand conservation constraints, not to signal redundancy.
Post-correction further decouples calibration from discrimination ($\Delta\text{Corr} = +0.067$, $p < 0.001$; Section~\ref{sec:decoupling}).

\item \textbf{Freely available corrections with direct deployment value.}

\gls{ntl} intensity (VIIRS DNB) and Proximity scores are publicly available at no cost; layered onto a static Voronoi pipeline, they reduce \gls{rmse} by $41\%$ and should be adopted as default corrections in any pipeline lacking a learned weighting component.
Method evaluation should report \gls{rmse} and correlation jointly, as the two metrics diverge under post-correction on learned representations.

\item \textbf{First ground-truth benchmark crossing weighting and integration mechanism.}

Prior reviews~\cite{Raventos2022comparison} compare models without metered ground truth; cross-national studies~\cite{Patil2024systematic} provide empirical breadth but do not vary the integration mechanism.
Using $1,891$ British primary substations, this study crosses demand weighting (static vs.\ \gls{gnn}) with integration mechanism (post-correction vs.\ prior-loss) in a controlled ground-truth evaluation---a two-dimensional comparison absent from existing literature.
Method rankings turn out to be mechanism-dependent rather than intrinsic to the model.
A German transfer check confirms directional replication and shows signs of degradation ($n = 13$, no statistical inference attempted) when the \gls{gnn} baseline already explains the dominant share of demand variance ($r^2 = 0.953$).
\end{enumerate}

The contribution is not a stronger \gls{gnn} but the identification of a failure mode in a reused upstream module of open-source energy system modeling.

The remainder of this paper is organized as follows:
Section~\ref{sec:related_work} reviews related work.
Sections~\ref{sec:methodology} and~\ref{sec:experimental_design} describe the methodology and the experimental design.
Section~\ref{sec:results} reports results.
Section~\ref{sec:discussion} discusses mechanisms, practical implications, and limitations.
Section~\ref{sec:conclusion} states conclusions.
\section{Related Work}
\label{sec:related_work}

Spatial disaggregation allocates regional aggregate quantities to finer spatial units.
The resolution at which these units are defined affects generation siting, transmission investment, and total system cost in energy optimization models \citep{Aryanpur2021review, Frysztacki2021strong, JalilVega2018effect, MartinezGordon2021review}.
The present paper does not address which resolution to adopt, but rather how to disaggregate aggregate data for fine-grained nodes once a resolution has been chosen.

\citet{Raventos2022comparison} compared the disaggregation procedures of eight German transmission network models along three dimensions: spatial partitioning, demand weighting, and auxiliary information integration.
Wide divergence appeared across all three dimensions, but the comparison remained inter-model because substation-level metered data were unavailable as ground truth.
\citet{Patil2024systematic} extended this taxonomy to climate action planning, sorting approaches into proxy-variable-based decomposition, machine learning, and hybrid methods, and provided cross-country empirical evidence by applying several of these methods to end-use sector consumption in Germany and Spain~\cite{Patil2025spatially}.
However, most of the methods mentioned in both papers lack validation using ground-truth data.

Along the demand-weighting dimension, early work merely relied on deterministic allocation rules.
\citet{Mei2016spatial} applied socio-demographic features to spatial estimation of electricity consumption using statistical learning.
Subsequent studies distributed electricity consumption and generation capacity across voltage levels using fixed proxy variables such as population, employment, and industrial output as weights \citep{Hulk2017allocation, Singh2015high, Koppl2017modeling}.
In prior work~\cite{Mu2026Improving}, we moved beyond this paradigm by showing that \gls{hgnn} can learn spatial weight functions from multimodal features, replacing rule-based demand weighting.

Along the spatial-partitioning dimension, \gls{vd} remains the default in both open-source frameworks~\cite{Horsch2018pypsa} and commercial tools; this paper fixes the partition to Voronoi and varies the remaining two axes.
Decomposition is inherently underdetermined: infinitely many allocations can reproduce a regional sum, which is why auxiliary information is needed to constrain the solution.
Adjacent fields supply direct evidence that the integration mechanism of domain knowledge is consequential, and that different mechanisms produce distinct effects.
\citet{Raventos2022comparison} identified post-correction as a procedural dimension but treated all corrections as a single category.
They neither distinguish inference-time correction from training-time guidance, nor compare their effects on disaggregation accuracy.
Neither static nor learning-based disaggregation methods have examined whether the same auxiliary data yields different results when it is used as an external overlay, an inference-time correction, or a training-time prior.

In physics-informed neural networks, \citet{Chen2024Physics-Informed} showed that encoding physical laws as loss-function penalties (soft constraints) enables the network to trade off physical consistency against data fit. 
The trade-off is governed by the loss coefficient: a low weight allows large violations, a high weight enforces near-exact compliance.
Encoding the same laws architecturally (hard constraints) operates through a different mechanism: it narrows the feasible solution space before optimization begins, enforcing physical consistency by construction rather than by gradient pressure.
Both mechanisms lead to distinct convergence behavior and generalization profiles, even when the underlying physical information is identical.

The same conclusion holds when domain priors are incorporated into graph-based and multi-task architectures.
\citet{Liang2022Knowledge} compared three entry points for domain priors in knowledge-guided graph models: input features (freely weighted by the model), graph structure (constraining which node interactions are representable), and auxiliary loss terms (creating competing gradient directions that reshape learned representations).
No single mechanism dominated across tasks.
\citet{Gong2019acomparison} reported an analogous finding in multi-task learning: hard parameter sharing forces shared low-level representations, while soft parameter sharing permits task-specific feature spaces connected by regularization; the choice between them, together with the loss weighting strategy, shifts the balance between positive transfer and negative interference.

In summary, based on the above-mentioned literature, the same information, entering a learning pipeline through different mechanisms, produces different model behavior.
Also, none of these results has been tested in spatial disaggregation.

Across the spatial disaggregation literature, the functional form through which auxiliary information enters the allocation is overwhelmingly multiplicative.
In weighted areal interpolation and dasymetric mapping, i.e.\ methods that redistribute zonal totals using auxiliary spatial data such as land cover, target-unit values are obtained by multiplying a base allocation by spatially varying weights derived from auxiliary data such as land use, building density, \gls{ntl}, and population grids, then re-normalizing to preserve the source-zone total \citep{Monteiro2019spatial, Sapena2022empiric, Qiu2022disaggregating}.
The same structure persists in hybrid pipelines that combine machine-learning density estimates with demand-conserving rescaling: the learned density field serves as a multiplicative weight, and a global scaling factor enforces conservation \citep{Georgati2024modeling, Monteiro2018hybrid}.
Even in emission inventory disaggregation, proxy intensities are combined multiplicatively before proportional allocation \citep{Patil2024systematic}.
The multiplicative form owes its prevalence to a structural convenience: it guarantees non-negativity and re-normalization, and automatically satisfies the demand conservation constraint.
Whether this convenience carries hidden costs when the base allocation is itself a learned model, rather than a uniform or rule-based prior, has not been examined.

\gls{ntl} imagery is among the most common proxy variables in spatial disaggregation.
\citet{Chen2022global} constructed a global $1\,\text{km} \times 1\,\text{km}$ gridded dataset of \gls{gdp} and electricity consumption from calibrated \gls{ntl} data and reported a strong positive correlation between \gls{ntl} intensity and electricity consumption at the global scale; \citet{Bhattarai2023remote} corroborated this at national and sub-national levels.
\gls{ntl} has also been used to disaggregate greenhouse gas emissions: \citet{Crippa2024insights} employed it as a spatial allocation proxy in the EDGAR\,v8.0 database, and \citet{Danylo2019high} combined building footprint, population density, and energy consumption data to produce high-resolution emission distributions for the residential sector.
The \gls{ntl}-electricity correlation is scale-dependent.
Meta-analyses report that predictive strength degrades at sub-national and sub-annual scales, where population, land use, and industrial structure become confounders \citep{Chen2022global, Bhattarai2023remote}.
Across these studies, \gls{ntl} is included in the allocation solely as a multiplicative correction factor applied to an initial estimate.
This form is natural when the initial estimate is rule-based or uniform, but a learning-based allocation model opens a second route: directly encoding the \gls{ntl} signal into the training objective.
Whether the two routes produce equivalent disaggregation accuracy has not been tested on ground-truth substation demand.

Substation proximity is a second auxiliary signal, motivated by the observation that spatially closer substations tend to carry more correlated loads \citep{Borges2020enhancing, Baran2005load}.
Proximity overlaps with \gls{ntl}: urban centers exhibit both high \gls{ntl} intensity and high substation density.
Whether combining both signals yields complementary gains or introduces redundant spatial correlation has not been tested.
Existing work uses geographic proximity or load density as features; to our knowledge, no study has performed ablation or information-theoretic analysis to disentangle the marginal contribution of each.



The preceding review exposes three research gaps:
\begin{enumerate}
    \item \citet{Raventos2022comparison} cataloged several auxiliary information integration practices, but no study has compared how these mechanisms affect disaggregation accuracy using ground-truth data.

    \item The physics-informed and multi-task learning literatures show that the entry point of domain knowledge into a learning pipeline is consequential~\citep{Chen2024Physics-Informed, Liang2022Knowledge, Gong2019acomparison}; this finding has not been tested in spatial load disaggregation.

    \item The partial overlap between \gls{ntl} intensity and Proximity has been noted but not quantified; whether their joint use is synergistic or antagonistic may itself depend on the integration mechanism.
\end{enumerate}

The paper addresses all three gaps through a controlled comparison of integration mechanisms and auxiliary signal combinations on a common disaggregation task validated against substation-level metered demand.
\section{Methodology}
\label{sec:methodology}

\subsection{Problem Definition}
\label{sec:problem}

Spatial load disaggregation distributes aggregate regional electricity demand to individual primary substations.
The inputs consist of $M$ administrative regions $\mathcal{R} = \{r_1, \ldots, r_M\}$ with known aggregate demand $D_r$; $N$ primary substations $\mathcal{T} = \{t_1, \ldots, t_N\}$ with geographic coordinates $(x_j, y_j)$ and metered peak demand $d_j^{\text{actual}}$; and a regular grid of $P$ spatial sample points (agents) $\mathcal{A} = \{a_1, \ldots, a_P\}$ covering the study area, each carrying a multi-source feature vector $\mathbf{f}_a$.

The objective is to recover a two-layer allocation mapping:
\begin{equation}
\hat{d}_j = \sum_{r \in \mathcal{R}} D_r \cdot \pi_{r \to j}, \quad \forall j \in \mathcal{T}
\label{eq:allocation}
\end{equation}
where $\pi_{r \to j}$ denotes the fraction of region $r$'s demand routed to substation $j$, subject to $\sum_j \pi_{r \to j} = 1$ and $\pi_{r \to j} \geq 0$ for all $r$.

An intermediate agent layer decomposes this transfer into two steps:
\begin{equation}
\pi_{r \to j} = \sum_{a \in \mathcal{A}_r} w_{ra} \cdot w_{aj}
\label{eq:transfer_decomposition}
\end{equation}
where $w_{ra}$ distributes demand from region $r$ to agent $a$, $w_{aj}$ routes demand from agent $a$ to substation $j$, and $\mathcal{A}_r$ is the set of agents within region $r$.

All methods evaluated here are expressed as a combination of three orthogonal design axes:
\begin{equation}
\begin{split}
\text{Method} = \;&\underbrace{\text{Spatial partitioning}}_{\text{Voronoi}}
\;\times\; \underbrace{\text{Demand weighting}}_{\text{Uniform / GPM / GNN}} \\
&\times\; \underbrace{\text{Auxiliary information integration}}_{\text{None / NTL / Prox / NTL+Prox}}
\end{split}
\label{eq:method_space}
\end{equation}

Eq.~\eqref{eq:method_space} assumes independence among the three axes; the experiments in Section~\ref{sec:asymmetry} test whether this holds.

For the third axis, auxiliary information enters the allocation through one of three mechanisms:
\begin{enumerate}
\item \textbf{\gls{gpm}-based post-correction}: multiplicative factors applied after \gls{gpm} demand shares are computed, adjusting the allocation before Voronoi aggregation.
\item \textbf{\gls{gnn}-based post-correction}: the same multiplicative factors, but applied to \gls{gnn} outputs after inference.
\item \textbf{\gls{gnn}-based prior loss}: additional loss terms injected during \gls{gnn} training that encode auxiliary spatial patterns, with no modification at inference time.
\end{enumerate}

\subsection{Spatial Partitioning}
\label{sec:voronoi}

Spatial partitioning determines which substation each agent feeds into.
Throughout this paper, partitioning follows Voronoi nearest-neighbor assignment, the default method in both open-source and commercial power system simulation tools.

Each agent $a$ is assigned to the geographically closest substation in projected Euclidean distance:
\begin{equation}
t^*(a) = \arg\min_{j \in \mathcal{T}_r} \| \mathbf{c}_a - \mathbf{c}_j \|_2
\label{eq:voronoi}
\end{equation}
where $\mathbf{c}_a$ and $\mathbf{c}_j$ are two-dimensional projected coordinates and $\mathcal{T}_r$ restricts candidates to substations within region $r$.
This produces a hard assignment: $w_{aj} = 1$ if $j = t^*(a)$ and zero otherwise.
The predicted demand at substation $j$ then equals the sum of agent demands within its Voronoi cell:
\begin{equation}
\hat{d}_j = \sum_{a:\, t^*(a) = j} d_a
\label{eq:voronoi_demand}
\end{equation}

\subsection{Demand Weighting}
\label{sec:demand_weighting}

Demand weighting determines the share of regional demand each agent carries.
Three methods are compared: uniform weighting as a baseline, and two published approaches, the Grid Point Model (GPM)~\cite{Mu2025Improving} and a heterogeneous Graph Neural Network (HGNN)~\cite{Mu2026Improving}.

\paragraph{Uniform Weighting}
\label{sec:uniform}
Regional demand is split equally among all agents in the region: $d_a = D_r / |\mathcal{A}_r|$.

\paragraph{Grid Point Model Weighting (GPM)}
\label{sec:gpm}
The \gls{gpm}~\cite{Mu2025Improving} assigns each agent to its dominant land-use class $k^*(a) = \arg\max_k\, p_a^{(k)}$ and weights it by the regional consumption share of that class, $\text{pct}_{r}^{(k^*)}$.
Weights are normalized per region to conserve $D_r$.
Residentially dominated agents, therefore, experience higher demand in regions where residential consumption accounts for a large fraction, and, conversely, other classes experience lower demand.

\paragraph{Graph Neural Network Weighting}
\label{sec:gnn_intro}
The \gls{gnn} method~\cite{Mu2026Improving} frames allocation as edge-weight prediction on a heterogeneous graph whose nodes represent sources (regions) and agents. 
A Heterogeneous Graph Transformer encoder produces node embeddings; allocation weights follow from a temperature-scaled softmax over learned edge costs, grouped by source:

\begin{equation}
w_{sa} = \frac{\exp(-c(s,a) / \tau)}{\sum_{a' \in \mathcal{N}(s)}
\exp(-c(s,a') / \tau)}, \quad
d_a^{\text{GNN}} = w_{s(a),a} \cdot D_{r(a)}
\label{eq:gnn_weights}
\end{equation}

where $c(s,a)$ is the learned cost of the edge connecting source $s$ to agent $a$, $\tau$ is a temperature hyperparameter, $\mathcal{N}(s)$ is the set of all agents connected to source $s$, $s(a)$ denotes the source node containing agent $a$, and $D_{r(a)}$ is the aggregate demand of the region to which $a$ belongs.
The resulting agent-level demand $d_a^{\text{GNN}}$ is the product of the allocation weight and the aggregate regional demand $D_{r(a)}$.

Training is weakly supervised via a land-use reconstruction loss:
\begin{equation}
\mathcal{L}_{\text{landuse}} = \frac{1}{|\mathcal{V}_s|}
\sum_{s} D_{\text{KL}}\!\left(\mathbf{q}_s^{\text{true}}
\;\big\|\; \text{norm}\!\left(\textstyle\sum_{a \in
\mathcal{N}(s)} w_{sa} \cdot \mathbf{M}_{a,:}\right)\right)
\label{eq:landuse_loss}
\end{equation}

where $\mathcal{V}_s$ is the set of source nodes, $\mathbf{q}_s^{\text{true}}$ is the observed consumption-share vector of source $s$ derived from regional economic statistics (detailed in Section~\ref{sec:experimental_design}), and $\mathbf{M}_{a,:}$ is the land-use proportion vector of agent $a$.
The loss penalizes deviations between the observed regional consumption structure and the structure reconstructed from weighted agent land-use vectors.

The proxy loss is necessary since most power networks lack substation-level demand labels; the \gls{gnn} therefore trains under weak supervision derived solely from region-level land-use consumption structures.

Three consequences of this self-supervised design are relevant to the integration analysis.
First, the \gls{gnn} trains without substation-level demand labels; supervision comes solely from the region-level land-use consumption structure.
Second, feature-level fusion, such as concatenating \gls{ntl} or Proximity values onto the 5-dimensional node feature vector, is ineffective under this training regime because no gradient signal incentivizes \gls{ntl}-informed allocation under this loss, so feature-level fusion lacks a learning channel.
Since Eq.(~\ref{eq:landuse_loss}) measures only land-use reconstruction quality, no gradient signal rewards \gls{ntl}-informed allocation; the network can discard features that do not reduce the land-use reconstruction error.
Third, the prior-loss mechanism (Section~\ref{sec:prior_loss}) is the only one that writes auxiliary spatial information into the training objective, generating a gradient signal aligned with the auxiliary pattern.
Post-correction (Section~\ref{sec:post_correction}) operates outside the training loop entirely and is therefore decoupled from the \gls{gnn}'s learned representation.


\subsection{Auxiliary Information}
\label{sec:auxiliary}

Two auxiliary information sources are examined: \gls{ntl} intensity and Proximity.
Both are entered as multiplicative correction factors.

\paragraph{\gls{ntl} Correction Factor}
\label{sec:ntl_factor}

\gls{ntl} radiance serves as a spatial proxy for human activity.
The correction factor takes the form:
\begin{equation}
\text{ntl\_factor}(a) = \frac{\log(1 + \text{ntl}(a) + \varepsilon)}{\log(1 + \tilde{m}_{\text{ntl}})}
\label{eq:ntl_factor}
\end{equation}
where $\text{ntl}(a)$ is the radiance at agent $a$ (nanoWatts/cm$^2$/sr), $\varepsilon$ is the 5th percentile of non-zero \gls{ntl} values among \gls{rci} agents, and $\tilde{m}_{\text{ntl}}$ is the median \gls{ntl} over \gls{rci} agents.

The floor $\varepsilon$ prevents dark areas (e.g., agricultural or forested land) from collapsing to zero while leaving the relative ordering among lit agents intact.
The logarithmic transform compresses the wide dynamic range of raw radiance, making the factor stable under moderate perturbations of $\varepsilon$.

\paragraph{Proximity Correction Factor}
\label{sec:prox_factor}

Primary substations tend to be sited near load centers, so an agent surrounded by many substations is more likely to lie in a high-demand zone.
The Proximity score sums inverse-distance contributions from all substations:
\begin{equation}
\text{prox}(a) = \sum_{j \in \mathcal{T}} d(a, j)^{-\gamma}
\label{eq:prox_score}
\end{equation}
where $d(a,j) = \max(\|\mathbf{c}_a - \mathbf{c}_j\|_2,\; 0.01\text{ km})$ is the projected Euclidean distance clamped at $10$\,m to prevent numerical overflow.

With $\gamma = 2$, contributions decay fast enough that only nearby substations carry appreciable weight.
The correction factor follows the same log-ratio form as the \gls{ntl} factor:
\begin{equation}
\text{prox\_factor}(a) = \frac{\log(1 + \text{prox}(a))}{\log(1 + \tilde{m}_{\text{prox}})}
\label{eq:prox_factor}
\end{equation}
with $\tilde{m}_{\text{prox}}$ the median Proximity score over \gls{rci} agents.

\subsection{Multiplicative Post-Correction}
\label{sec:post_correction}

Both the \gls{gpm} and \gls{gnn} pipelines can incorporate auxiliary information through the same mechanism: agent-level demands are multiplied by a correction factor and re-normalized within each sub-region to conserve $D_r$.
Given a base demand $d_a^{\text{base}}$ and correction factor $f(a)$, the corrected demand is:
\begin{equation}
d_a^{\text{corr}} = D_r \cdot \frac{d_a^{\text{base}} \cdot f(a)}{\sum_{a' \in \mathcal{A}_r^{\text{sub}}} d_{a'}^{\text{base}} \cdot f(a')}
\label{eq:post_correction}
\end{equation}
where $f(a)$ is $\mathrm{ntl\_factor}(a)$, $\mathrm{prox\_factor}(a)$, or both applied sequentially.

For \gls{gnn}-based methods, the base allocation $w_{sa}$ is already normalized by the per-source softmax, see Eq.~\ref{eq:gnn_weights}.
Eq.~\eqref{eq:post_correction} therefore introduces a second normalization step.
This is a deliberate choice that guarantees demand conservation at the sub-regional level, but it distorts the relative weights learned by the \gls{gnn}.
Specifically, the softmax allocation encodes the network's belief about inter-agent demand ratios; the re-normalization in Eq.~\eqref{eq:post_correction} overrides these ratios in proportion to the variance of $f(a)$ within each sub-region.
High-variance correction factors (e.g., \gls{ntl} in regions with mixed urban and rural land) produce larger distortions than low-variance ones.

\subsection{Prior Loss}
\label{sec:prior_loss}

The alternative to post-correction is to inject auxiliary information during \gls{gnn} training rather than after inference.
We term this the \emph{prior-loss} mechanism because auxiliary information is incorporated prior to inference during training, in contrast to post-correction, which operates after inference; the term does not imply a Bayesian prior over model parameters.
A prior loss term converts the \gls{ntl} or Proximity distribution into a target weight distribution and penalizes deviation via forward KL divergence.

For each source $s$, the target distribution over \gls{rci} agents is:
\begin{equation}
q_{s,a} = \frac{\log(1 + v(a))}{\sum_{a' \in \mathcal{A}_s \cap \mathrm{RCI}} \log(1 + v(a'))}
\label{eq:prior_target}
\end{equation}
where $v(a)$ is $\mathrm{ntl}(a)$ or $\mathrm{prox}(a)$.
The prior loss is:
\begin{equation}
\mathcal{L}_{\text{prior}} = \frac{1}{|\mathcal{V}_s|} \sum_{s} \sum_{a \in \mathcal{A}_s} q_{s,a} \log\!\left(\frac{q_{s,a}}{w_{sa}}\right)
\label{eq:prior_loss}
\end{equation}

Forward KL ($\mathrm{KL}(q \| w)$) is chosen over its reverse ($\mathrm{KL}(w \| q)$) for a practical reason:
It penalizes the network for under-allocating demand to agents that the auxiliary source marks as active, but does not penalize surplus allocation to agents where the auxiliary signal is weak.
In load disaggregation terms, the network must respect high-activity locations flagged by \gls{ntl} or Proximity, while remaining free to assign demand elsewhere based on land-use or other learned features.
Reverse KL would have the opposite effect, concentrating demand on a few dominant agents and suppressing allocation elsewhere.
This is undesirable when the auxiliary source is noisy or incomplete.

The total training objective is:
\begin{equation}
\mathcal{L}_{\text{total}} = \mathcal{L}_{\text{landuse}} + \lambda_{\text{ntl}}\,\mathcal{L}_{\text{ntl-prior}} + \lambda_{\text{prox}}\,\mathcal{L}_{\text{prox-prior}}
\label{eq:total_loss}
\end{equation}

Unlike post-correction, the prior-loss mechanism lets the network itself decide how much of the auxiliary signal to absorb, without introducing a second normalization at inference time.
The trade-off is retraining: each prior configuration requires a full training run, whereas post-correction can be applied to any pre-trained model without additional computation.

\subsection{Summary of Evaluated Methods}
\label{sec:method_matrix}

Tab.~\ref{tab:methods} lists all $15$ methods evaluated in this paper, classified along the three design axes, demand weighting, and auxiliary information integration are systematically varied.
The spatial partitioning axis is fixed to Voronoi, the current industry standard; while alternative partitioning methods exist, fixing this axis isolates the effects of the other two.

\begin{table}[htbp]
\centering
\caption{Method matrix.
All methods use Voronoi spatial partitioning.
N = \gls{ntl}, P = Proximity ($\gamma{=}2$), NP = both.
Within each group, methods are ordered by increasing auxiliary information.}
\label{tab:methods}
\setlength{\tabcolsep}{2pt}
\renewcommand{\arraystretch}{1.1}
\small
\begin{tabular}{ll cccc cccc ccccccc}
\toprule
& & \multicolumn{4}{c}{Uniform base} & \multicolumn{4}{c}{\gls{gpm} base} & \multicolumn{7}{c}{\gls{gnn} base} \\
\cmidrule(lr){3-6} \cmidrule(lr){7-10} \cmidrule(lr){11-17}
& &
\rotatebox{90}{Uni} &
\rotatebox{90}{UniP} &
\rotatebox{90}{UniN} &
\rotatebox{90}{UniNP} &
\rotatebox{90}{\gls{gpm}} &
\rotatebox{90}{GPMpostP} &
\rotatebox{90}{GPMpostN} &
\rotatebox{90}{GPMpostNP} &
\rotatebox{90}{\gls{gnn}} &
\rotatebox{90}{GNNpostP} &
\rotatebox{90}{GNNpostN} &
\rotatebox{90}{GNNpostNP} &
\rotatebox{90}{GNNpriorP} &
\rotatebox{90}{GNNpriorN} &
\rotatebox{90}{GNNpriorNP} \\
\midrule
\multirow{3}{*}{\rotatebox{90}{\footnotesize Weight}}
& Uniform & \yes & \yes & \yes & \yes &      &      &      &      &     &     &     &     &     &     &     \\
& \gls{gpm}     &      &      &      &      & \yes & \yes & \yes & \yes &     &     &     &     &     &     &     \\
& \gls{gnn}     &      &      &      &      &      &      &      &      & \yes& \yes& \yes& \yes& \yes& \yes& \yes\\
\midrule
\multirow{2}{*}{\rotatebox{90}{\footnotesize Aux}}
& \gls{ntl}     &      &      & \yes & \yes &      &      & \yes & \yes &     &     & \yes& \yes&     & \yes& \yes\\
& Prox    &      & \yes &      & \yes &      & \yes &      & \yes &     & \yes&     & \yes& \yes&     & \yes\\
\midrule
\multirow{2}{*}{\rotatebox{90}{\footnotesize Path}}
& Post.  &  & \yes & \yes & \yes &      & \yes & \yes & \yes &     & \yes& \yes& \yes&     &     &     \\
& Prior.  &  &      &      &      &      &      &      &      &     &     &     &     & \yes& \yes& \yes\\
\bottomrule
\end{tabular}
\end{table}
\section{Experimental Design}
\label{sec:experimental_design}

The primary study area comprises $16$ \gls{itl}-2 zones in Great Britain, each populated with approximately $50{,}000$ grid points that serve as the spatial units (agents) for demand allocation.
Each agent is assigned to a substation via \gls{vd}.
To test whether the core findings transfer across networks, a secondary case study is drawn from the B\"{o}rde district in Sachsen-Anhalt, Germany, a NUTS 3 administrative unit containing $13$ primary substations and $34$ municipalities.

\subsection{Data Sources and Processing}
\label{sec:data}

Tab.~\ref{tab:data_sources} summarizes the datasets used in both study areas.
The German case follows the same processing pipeline; only the projected coordinate reference system differs (EPSG:$25832$ for Germany versus EPSG:$27700$ for Britain).

\begin{table*}[b]
\centering
\caption{Data sources for the British and German study areas.}
\label{tab:data_sources}
\begin{tabular}{llll}
\toprule
\textbf{Data} & \textbf{Resolution} & \textbf{Role} \\
\midrule
Administrative boundaries~\cite{uk_itl, eurostat_gisco_nuts} & --- & Region delineation \\
Substation locations \& demand~\cite{Zhou2024Dataset, regiocom2026avacon} & Point & Ground truth \\
Land use~\cite{openstreetmap_contributors_planet_2017} & $\sim$100 tags $\to$ 5 classes & \gls{gpm} weights; \gls{gnn} features \\
Nighttime Light (NTL)~\cite{elvidge2017viirs} & $\sim$500\,m & \gls{ntl} correction factor \\
Economic output~\cite{uk_gva, uk_population, eurostat_nama10r3gva} & \gls{itl}-3 & \gls{gnn} reconstruction loss \\
\bottomrule
\end{tabular}
\end{table*}

\paragraph{Land-use classification}
OpenStreetMap tags are aggregated into five functional classes: residential, commercial, industrial, agricultural, and other.
For each agent~$a$, a land-use proportion vector $\mathbf{p}_a = (p_a^{\mathrm{res}},\, p_a^{\mathrm{com}},\, p_a^{\mathrm{ind}},\, p_a^{\mathrm{agr}},\, p_a^{\mathrm{oth}}) \in \mathbb{R}^5$ is obtained by spatial join.
This vector feeds both the \gls{gpm} weighting rules and the \gls{gnn} node feature matrix.

\paragraph{Nighttime light}
\gls{ntl} radiance is taken from the NOAA VIIRS Day/Night Band (DNB) monthly composites.
The median of all cloud-free observations between January~2021 and December~2022 is used, matching the 2021--2022 reporting year of the British substation demand data~\cite{Zhou2024Dataset}.
The \texttt{avg\_rad} band records mean radiance in nanoWatts\,cm$^{-2}$\,sr$^{-1}$; the observed dynamic range across the British study area spans 0--267. The native pixel size is approximately $500~\mathrm{m}$.
Each agent receives an \gls{ntl} value by bilinear interpolation; agent-level values within a Voronoi cell are then aggregated into the \gls{ntl} correction factor defined in Section~\ref{sec:methodology}.

\subsection{Cross-Validation and Evaluation Protocol}
\label{sec:cv_protocol}

\begin{figure}[htbp]
  \centering
  \includegraphics[width=\columnwidth]{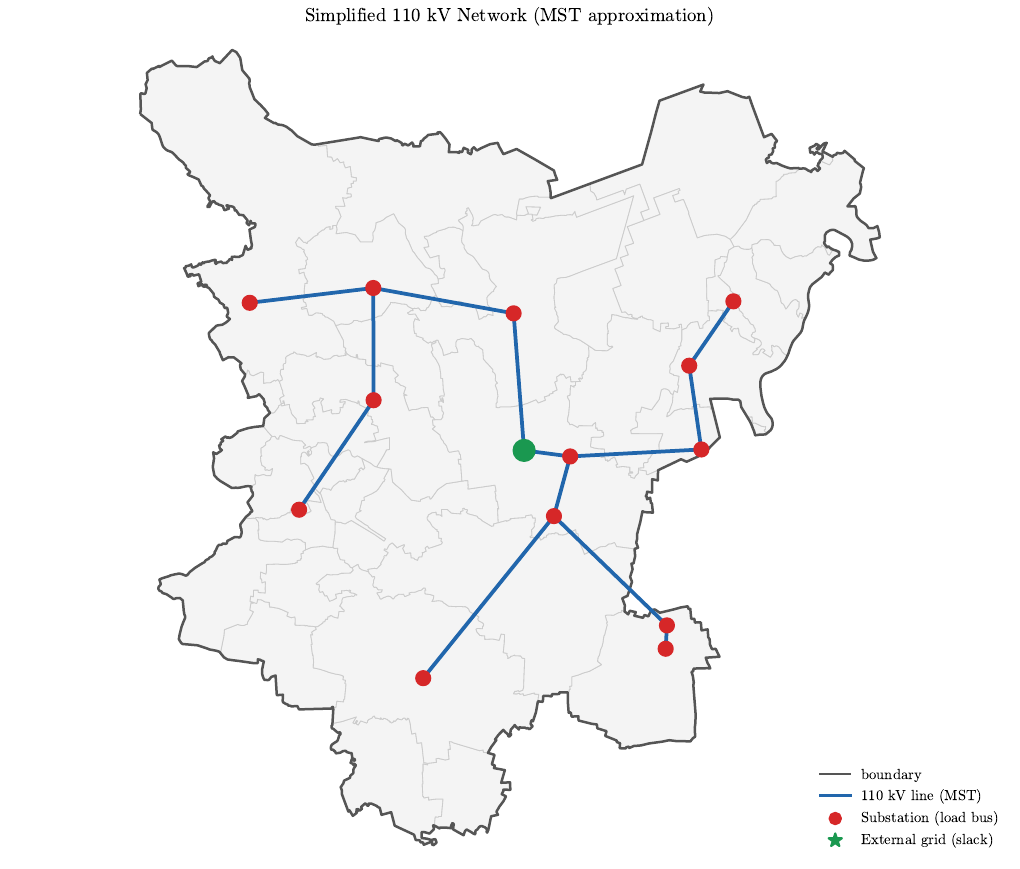}
  \caption{Simplified 110\,kV network for the B\"{o}rde district used for downstream power-flow validation.
  Red circles represent the 13 primary substations, each annotated with its measured peak demand; the green star marks the slack bus, positioned at the centroid, representing the external grid connection. Blue edges are the 13 minimum spanning tree connections, labeled with line lengths in kilometers. Grey polygons show the 34\, Gemeinde boundaries that define the demand aggregation units.}
  \label{fig:boerde_network}
\end{figure}

\subsubsection{British main case}

A four-fold spatial cross-validation scheme is used.
The $16$ \gls{itl}-2 regions are split into four groups of four; in each fold, three groups ($12$~regions) form the training set, and the remaining group (4~regions) is held out for testing.
This procedure runs under three random seeds ($42$, $123$, $456$) controlling both parameter initialization and fold assignment, producing $4 \times 3 = 12$ evaluation runs per \gls{gnn}-based method.

Although spatial autocorrelation exists between adjacent \gls{itl}-2 regions, each region is modeled as a separate graph: source and agent nodes belong to exactly one region, and no edge or message passing crosses region boundaries.
Encoder parameters are shared across all training-set graphs within a fold, but the test-region graph is withheld during training, so no substation-level information from the test region can leak into the learned weights.

All methods are reported as the $16$-region mean $\pm$ inter-region standard deviation.
For \gls{gnn}-based methods, per-region metric values are first averaged over the three seeds before regional aggregation; the reported $\pm$ therefore reflects spatial heterogeneity, not training randomness.
Static methods (Uniform, \gls{gpm}, and their post-corrected variants) are deterministic and require no seed averaging.

\subsubsection{German validation case}
\label{subsec:germany_setup}

All $13$ substations and $34$ municipalities participate in a single train-and-evaluate pass using the same three seeds.
Results are reported as the three-seed mean $\pm$ standard deviation.
Given the small sample size, no statistical inference is attempted.

To address the circularity inherent in train-on-test evaluation, a jackknife leave-one-out (LOO) analysis is performed: each of the $13$ substations is removed in turn, and the correlation among the remaining $12$ is recomputed, yielding $13$ LOO estimates per seed.

To assess whether RMSE differences translate into engineering consequences, a 14-bus simplified 110\,kV network is constructed from the substation coordinates of the 13 B\"{o}rde primary substations. 
A minimum spanning tree, computed over the pairwise geodesic distance matrix of the 13 substations, defines 13 edges that connect all load nodes into a radial topology. 
Each edge is parameterized with the standard German 110\,kV overhead-conductor type 149-AL1/24-ST1A. 
A fourteenth bus, representing the connection to the external high-voltage transmission grid, is placed at the geographic centroid of the 13 substation coordinates and designated as the slack bus at $v = 1.02\,\text{p.u.}$ AC power flow is solved with \textit{pandapower}~\cite{thurner2018pandapower}. 
The minimum-spanning-tree topology and the centroid-positioned slack bus are spatial approximations that do not reproduce the actual Avacon~Netz 110\,kV grid; all quantities derived from this model quantify relative differences among disaggregation methods under identical network assumptions, and absolute loading values should not be read as statements about the real B\"{o}rde network. Fig.~\ref{fig:boerde_network} shows the resulting topology overlaid on the 34 municipality boundaries.

\subsubsection{Evaluation metrics}

Three metrics are computed at the substation level within each region~$r$.
The root-mean-square error,
\begin{equation}
\mathrm{RMSE}_r = \sqrt{\frac{1}{|\mathcal{T}_r|} \sum_{j \in \mathcal{T}_r} \bigl(\hat{d}_j - d_j^{\mathrm{actual}}\bigr)^2}\,,
\label{eq:rmse}
\end{equation}
measures the absolute accuracy of disaggregated demand.
The mean absolute error,
\begin{equation}
\mathrm{MAE}_r = \frac{1}{|\mathcal{T}_r|} \sum_{j \in \mathcal{T}_r} \bigl|\hat{d}_j - d_j^{\mathrm{actual}}\bigr|\,,
\label{eq:mae}
\end{equation}
gives the average absolute deviation and is less sensitive to outliers than \gls{rmse}.
The Pearson correlation coefficient,
\begin{equation}
\mathrm{Corr}_r = \frac{\displaystyle\sum_{j} \bigl(\hat{d}_j - \bar{\hat{d}}\bigr)\bigl(d_j^{\mathrm{actual}} - \bar{d}^{\mathrm{actual}}\bigr)}{\sqrt{\displaystyle\sum_{j} \bigl(\hat{d}_j - \bar{\hat{d}}\bigr)^2}\;\sqrt{\displaystyle\sum_{j} \bigl(d_j^{\mathrm{actual}} - \bar{d}^{\mathrm{actual}}\bigr)^2}}\,,
\label{eq:corr}
\end{equation}
captures the linear association between predicted and observed substation demands.

\begin{table*}[htbp]
\centering
\caption{Substation-level performance on the British study area (Voronoi partition, $16$-region mean $\pm$ std).
All methods report inter-region variability ($n = 16$ regions); \gls{gnn} method values are seed-averaged over three random initializations prior to aggregation.
Best overall in bold; best \gls{gnn} underlined.}

\label{tab:main}
\small
\begin{tabular}{llccc}
\toprule
\textbf{Method} & \textbf{Type} & \textbf{RMSE} & \textbf{MAE} & \textbf{Corr} \\
\midrule
\multicolumn{5}{l}{\emph{Static methods (inter-region std)}} \\
Uni        & Uniform baseline          & $13.10 \pm 3.70$ & $9.23 \pm 2.21$ & $0.009 \pm 0.178$ \\
UniN       & Uniform + \gls{ntl}             & $11.16 \pm 3.29$ & $7.96 \pm 2.02$ & $0.094 \pm 0.190$ \\
UniP       & Uniform + Prox            & $8.98 \pm 2.96$  & $6.71 \pm 1.89$ & $0.119 \pm 0.194$ \\
UniNP      & Uniform + \gls{ntl} + Prox      & $7.45 \pm 2.79$  & $5.50 \pm 1.68$ & $0.319 \pm 0.194$ \\
\gls{gpm}        & \gls{gpm} weighting             & $12.39 \pm 3.36$ & $8.74 \pm 2.11$ & $0.035 \pm 0.199$ \\
GPMpostN   & \gls{gpm} + \gls{ntl}                 & $10.54 \pm 2.80$ & $7.56 \pm 1.84$ & $0.128 \pm 0.208$ \\
GPMpostP   & \gls{gpm} + Prox                & $8.55 \pm 2.50$  & $6.45 \pm 1.72$ & $0.152 \pm 0.179$ \\
\textbf{GPMpostNP} & \textbf{\gls{gpm} + \gls{ntl} + Prox} & $\mathbf{7.31 \pm 2.40}$ & $\mathbf{5.44 \pm 1.50}$ & $\mathbf{0.332 \pm 0.173}$ \\
\midrule
\multicolumn{5}{l}{\emph{\gls{gnn} methods (seed-averaged)}} \\
\gls{gnn}        & \gls{gnn} baseline              & $9.27 \pm 2.61$  & $6.67 \pm 1.76$ & $0.298 \pm 0.201$ \\
GNNpostN   & \gls{gnn} + \gls{ntl} post-corr.      & $9.45 \pm 2.55$  & $6.58 \pm 1.64$ & $0.365 \pm 0.187$ \\
GNNpostP   & \gls{gnn} + Prox post-corr.     & $9.42 \pm 2.33$  & $6.59 \pm 1.52$ & $0.349 \pm 0.166$ \\
GNNpostNP  & \gls{gnn} + \gls{ntl} + Prox post-corr. & $11.20 \pm 2.64$ & $7.64 \pm 1.66$ & $0.348 \pm 0.160$ \\
GNNpriorN  & \gls{gnn} + \gls{ntl} prior           & $9.19 \pm 3.02$  & $6.65 \pm 1.96$ & $0.335 \pm 0.173$ \\
GNNpriorP  & \gls{gnn} + Prox prior          & $9.14 \pm 2.89$  & $6.60 \pm 1.88$ & $0.335 \pm 0.176$ \\
\underline{GNNpriorNP} & \underline{\gls{gnn} + \gls{ntl} + Prox prior} & $\underline{9.03 \pm 2.88}$ & $\underline{6.51 \pm 1.86}$ & $\underline{0.342 \pm 0.179}$ \\
\bottomrule
\end{tabular}
\end{table*}

\section{Results}
\label{sec:results}

The results are organized into seven subsections.
Section~\ref{sec:main_results} compares aggregate performance and statistical significance across all $15$~methods.
The \gls{rmse}--\gls{corr} decoupling phenomenon is documented in Section~\ref{sec:decoupling}; Section~\ref{sec:asymmetry} then quantifies how synergistic (error-reducing) and antagonistic (error-amplifying) interaction effects differ across integration mechanisms.
Section~\ref{sec:mechanism_isolation} isolates the structural cause of the antagonism through controlled mechanism experiments.
Section~\ref{sec:sensitivity} checks these patterns against intensity sweeps to rule out hyperparameter artifacts.
The remaining two subsections shift focus from method behavior to context dependence: Section~\ref{sec:heterogeneity} asks whether regional characteristics modulate the observed effects, and Section~\ref{sec:germany} tests cross-network transferability on the German B\"{o}rde case.

\subsection{Aggregate Performance and Statistical Significance}
\label{sec:main_results}

Tab.~\ref{tab:main} reports the \gls{rmse}, \gls{mae}, and \gls{corr} for all $15$~methods under the Voronoi partition.
All values are the mean across $16$~\gls{itl}-2 regions $\pm$ inter-region standard deviation; \gls{gnn} method values are seed-averaged over three random initializations prior to regional aggregation.

GPMpostNP ($7.31$) achieves the lowest \gls{rmse}, followed closely by UniNP ($7.45$; +$0.14$). 
A gap of $1.10$ separates this pair from the next cluster: GPMpostP ($8.55$), UniP ($8.98$), and the three prior-loss GNN variants ($9.03$–$9.19$), which span a narrow $0.16$ range. 
The uncorrected GNN ($9.27$) and both single post-corrections ($9.42$, $9.45$) fall within $0.2$ of each other. 
GNNpostNP ($11.20$) stands apart, its RMSE approaching the GPM baseline ($12.39$).

Tab.~\ref{tab:significance} reports the Wilcoxon signed-rank test results for nine planned comparisons, with $p$-values adjusted by the Holm--Bonferroni method.
Each test uses $16$~paired regions; the reported $p$-value is the median across three seeds.

\begin{table}[htbp]
\centering
\caption{Pairwise statistical tests (Wilcoxon signed-rank, Holm--Bonferroni corrected).
Positive $\Delta$\gls{rmse} indicates deterioration; negative indicates improvement.
Tests on \gls{rmse}, \gls{mae}, and \gls{corr} are conducted separately.}
\label{tab:significance}
\small
\begin{tabular}{clccc}
\toprule
\textbf{\#} & \textbf{Comparison} & \textbf{$\Delta$\gls{rmse}} & \textbf{Holm $p$} & \textbf{Sig.} \\
\midrule
1 & \gls{gnn} vs \gls{gpm}                  & $-3.12$ & $6.1 \times 10^{-4}$  & \checkmark \\
2 & GNNpostP vs \gls{gnn}             & $+0.15$ & $1.0$                  & --- \\
3 & GNNpostN vs \gls{gnn}             & $+0.19$ & $1.0$                  & --- \\
4 & GNNpostNP vs GNNpostP       & $+1.78$ & $2.1 \times 10^{-4}$  & \checkmark \\
5 & GPMpostNP vs GPMpostN       & $-3.23$ & $3.1 \times 10^{-4}$  & \checkmark \\
6 & GPMpostP vs \gls{gpm}             & $-3.85$ & $2.8 \times 10^{-4}$  & \checkmark \\
7 & GPMpostNP vs GPMpostP       & $-1.24$ & $2.8 \times 10^{-4}$  & \checkmark \\
8 & GNNpriorN vs \gls{gnn}            & $-0.08$ & $1.0$                  & --- \\
9 & GNNpostP vs GPMpostNP       & $+2.11$ & $2.1 \times 10^{-4}$  & \checkmark \\
\bottomrule
\end{tabular}
\end{table}

Six of the nine comparisons are statistically significant.
They point to three conclusions.

\gls{gnn} significantly outperforms \gls{gpm} (comparison~\#1: $\Delta\text{RMSE} = -3.12$, $p < 0.001$), which confirms that end-to-end learning on the same five-dimensional land-use features can replace hand-crafted weighting rules.

The static correction chain, by contrast, exhibits layer-wise synergy.
Proximity added to \gls{gpm} yields $\Delta\text{RMSE} = -3.85$ (\#6); \gls{ntl} added on top of \gls{gpm}+Proximity yields a further $-1.24$ (\#7); Proximity added on top of \gls{gpm}+\gls{ntl} yields $-3.23$ (\#5).
All three comparisons are significant at $p < 0.001$, so each correction layer contributes information that the others do not already capture.

For the \gls{gnn} base, the pattern reverses.
The combined post-correction is significantly antagonistic: GNNpostNP versus GNNpostP gives $\Delta\text{RMSE} = +1.78$ ($p < 0.001$, \#4).
The static optimum GPMpostNP also significantly outperforms the best \gls{gnn} post-correction GNNpostP, with $\Delta\text{RMSE} = +2.11$ ($p < 0.001$, \#9).

The three non-significant comparisons reinforce rather than weaken this reading.
Single post-corrections on \gls{gnn} (\#2, \#3) show consistent \gls{rmse} deterioration ($+0.15$, $+0.19$) that does not reach significance.
The prior-loss \gls{ntl} variant (\#8) points in the beneficial direction ($-0.08$) but does not reach significance ($p = 0.67$).
On the \gls{rmse} metric, then, auxiliary information grafted onto a trained \gls{gnn} provides no significant benefit regardless of mechanism; the prior-loss route is simply the only one that avoids outright degradation.

\subsection{RMSE--Corr Decoupling}
\label{sec:decoupling}

Applying the same statistical tests to the \gls{corr} metric exposes a decoupling between calibration (absolute accuracy, measured by \gls{rmse}) and discrimination (rank ordering, measured by \gls{corr}) that is invisible to either metric alone.
Tab.~\ref{tab:decoupling} contrasts the \gls{rmse} and \gls{corr} significance levels for four representative comparisons.
For GNNpostN relative to \gls{gnn}, \gls{rmse} exhibits a non-significant increase ($+0.19$, $p = 1.0$), whereas \gls{corr} improves significantly ($+0.067$, from $0.298$ to $0.365$, $p = 5.5 \times 10^{-4}$).
GNNpostP follows the same pattern: \gls{rmse} increases by $+0.15$ while \gls{corr} rises by $+0.051$.

\begin{table}[htbp]
\centering
\caption{\gls{rmse}--\gls{corr} decoupling: the same comparison can be significant on one metric and non-significant on the other.
Holm--Bonferroni corrected $p$-values; three-seed median.}
\label{tab:decoupling}
\small
\begin{tabular}{lcccc}
\toprule
 & \multicolumn{2}{c}{\textbf{\gls{rmse}}} & \multicolumn{2}{c}{\textbf{\gls{corr}}} \\
\cmidrule(lr){2-3} \cmidrule(lr){4-5}
\textbf{Comparison} & $\Delta$ & Sig. & $\Delta$ & Sig. \\
\midrule
GNNpostN vs \gls{gnn}      & $+0.19$  & ---        & $+0.067$ & \checkmark \\
GNNpostP vs \gls{gnn}      & $+0.15$  & ---        & $+0.051$ & ---        \\
GNNpostNP vs GNNpostP & $+1.78$ & \checkmark & $-0.001$ & ---        \\
GPMpostNP vs GPMpostN & $-3.23$ & \checkmark & $+0.204$ & \checkmark \\
\bottomrule
\end{tabular}
\end{table}

The combined post-correction (GNNpostNP) further sharpens the pattern.
Its \gls{corr} of $0.348$ is statistically indistinguishable from that of GNNpostP ($0.349$; $p = 0.98$), yet \gls{rmse} deteriorates sharply to $11.20$.
The third row of Tab.~\ref{tab:decoupling} confirms this asymmetry: the incremental \gls{ntl} factor is significant on \gls{rmse} (antagonistic) but non-significant on \gls{corr} (no ranking gain).

The decoupling does not arise in the static pipeline.
GPMpostNP achieves the highest \gls{corr} ($0.332$) and the lowest \gls{rmse} ($7.31$) simultaneously, as shown in the last row of Tab.~\ref{tab:decoupling}, where both metrics are significant and improving.
On the \gls{gpm} base (\gls{corr} $= 0.035$), correction layers improve both calibration and discrimination in tandem.
On the \gls{gnn} base (\gls{corr} $= 0.298$), the same corrections improve discrimination while worsening calibration--- a reversal of the pattern.

\subsection{Synergy--Antagonism Asymmetry}
\label{sec:asymmetry}

To quantify how identical auxiliary information behaves under different integration mechanisms, the marginal \gls{rmse} change $\Delta_I^B = \text{RMSE}(B + I) - \text{RMSE}(B)$ is computed for each information component~$I$ on each base~$B$; negative values indicate improvement.
A positive marginal effect is termed \textit{antagonistic} when it is statistically significant ($p < 0.05$, Holm--Bonferroni corrected); a positive but non-significant marginal effect is classified as \textit{directionally adverse}.

\begin{table*}[htbp]
\centering
\caption{Independent marginal \gls{rmse} effect of each auxiliary component, measured from the unmodified base: $\Delta = \text{RMSE}(\text{base} + I) - \text{RMSE}(\text{base})$.
Negative values denote improvement; positive values denote deterioration.
Each row reports the effect of adding that component alone; the \gls{ntl}+Prox row adds both simultaneously.
Significance assessed via Wilcoxon signed-rank test with Holm--Bonferroni correction.}
\label{tab:asymmetry}
\begin{tabular}{lcccc}
\toprule
\textbf{Component} & \textbf{Uniform base} & \textbf{\gls{gpm} base} & \textbf{\gls{gnn} post-corr.} & \textbf{\gls{gnn} prior loss} \\
\midrule
\gls{ntl}           & $-1.94$ ($-14.8\%$)   & $-1.85$ ($-15.0\%$)  & $+0.19$ ($+2.0\%$)     & $-0.08$ ($-0.9\%$) \\
Proximity     & $-4.12$ ($-31.5\%$)   & $-3.85$ ($-31.0\%$)  & $+0.15$ ($+1.6\%$)     & $-0.13$ ($-1.4\%$) \\
\gls{ntl}+Prox      & $\mathbf{-5.65}$ ($-43.1\%$) & $\mathbf{-5.08}$ ($-41.0\%$) & $\mathbf{+1.93}$ ($+20.8\%$) & $-0.24$ ($-2.6\%$) \\
\bottomrule
\end{tabular}
\end{table*}

The asymmetry across mechanisms is pronounced.
On the static bases, \gls{ntl} and Proximity are approximately additive: the combined \gls{gpm} effect ($-5.08$) accounts for $89\%$ of the sum of independent marginal effects ($-1.85 + (-3.85) = -5.70$), and the corresponding ratio on Uniform is $93\%$ ($-5.65$ versus $-6.06$).
The slight sub-additivity reflects partial overlap between \gls{ntl} and Proximity in urbanized areas, though all $16$ regions improve in every case ($p < 0.001$).

On the \gls{gnn} post-correction route, single-factor corrections are directionally adverse but statistically non-significant (NTL alone: $+0.19$, $p = 1.0$; Proximity alone: $+0.15$, $p = 1.0$).
As documented in Section~\ref{sec:decoupling}, these corrections simultaneously produce a significant \gls{corr} improvement ($+0.067$), marking them as instances of metric decoupling rather than antagonism.
Combining \gls{ntl} and Proximity crosses the significance threshold and constitutes true antagonism: $+1.93$~\gls{rmse}, $+20.8\%$, $p < 0.001$.
This combined effect is also super-additive---the deterioration ($+1.93$) far exceeds the sum of the individual directionally adverse effects ($+0.19 + 0.15 = +0.34$)---and the incremental contribution of adding \gls{ntl} on top of Proximity correction is itself significant ($\Delta\text{RMSE} = +1.78$, $p < 0.001$, Tab.~\ref{tab:significance} \#4).

On the prior-loss route, the combined effect ($-0.24$~\gls{rmse}, $-2.6\%$) points in the synergistic direction but does not reach significance ($p = 0.67$), and the improvement remains small.

\paragraph{Auxiliary-signal correlation.}
The \gls{ntl} and Proximity correction factors are moderately correlated across $122{,}441$ \gls{rci} agents (Pearson $r = 0.350$, Spearman $\rho = 0.481$, $p < 0.001$).
Both factors encode aspects of urbanization intensity (\gls{ntl} through nocturnal radiance, Proximity through substation siting density), and their shared variance concentrates in urban cores ($\rho = 0.685$ in the densest region versus $\rho = 0.241$ in the most rural).
Section~\ref{sec:mechanism_isolation} examines whether this correlation is a primary driver of the observed antagonism or a secondary factor.

\subsection{Mechanism Isolation}
\label{sec:mechanism_isolation}

To disentangle the causes of post-correction antagonism on the \gls{gnn} base, three diagnostic experiments are conducted.
Each follows the same $3 \times 4 = 12$ seed--fold evaluation protocol used in the main experiments.
Tab.~\ref{tab:mechanism_isolation} collects the full numerical results for all three experiments and their single-factor variants.

\paragraph{Experiment~1: Removing re-normalization}
If the antagonism were an artifact of double-normalization, where the \gls{gnn}'s per-source softmax is followed by the post-correction re-normalization in Eq.~\ref{eq:post_correction}, then removing the denominator in Eq.~\ref{eq:post_correction} and applying raw multiplicative scaling should eliminate it.
The outcome is the opposite (Tab.~\ref{tab:mechanism_isolation}): without re-normalization, \gls{rmse} rises from $11.20$ to $17.53$ ($+56.5\%$), because demand conservation is violated and allocations lose calibration.
Double-normalization is therefore not the cause of antagonism; re-normalization is a structural necessity.

\paragraph{Experiment~2: Random-noise post-correction}
If the antagonism stems from signal redundancy, i.e.\ the \gls{gnn} having already encoded \gls{ntl} and Proximity information so that post-correction double-counts it, then replacing the real correction factors with spatially uncorrelated random noise should eliminate the effect.
Random-noise multiplicative post-correction with re-normalization yields \gls{rmse} $= 9.59 \pm 2.68$ (Tab.~\ref{tab:mechanism_isolation}), only marginally above the uncorrected \gls{gnn} baseline ($9.27$).
This is far below the real \gls{ntl}+Proximity multiplicative correction ($11.20$), indicating that the spatial signal itself, not the normalization structure, drives the antagonism.

\paragraph{Experiment~3: Additive post-correction}
If the antagonism is a structural consequence of the multiplicative correction form under conservation constraints, where multiplicative factors disproportionately amplify high-demand areas in a zero-sum redistribution, then switching to an additive form should remove it.
Additive \gls{ntl}+Proximity post-correction yields \gls{rmse} $= 7.34 \pm 2.14$ and \gls{corr} $= 0.448 \pm 0.188$ (Tab.~\ref{tab:mechanism_isolation}), a large improvement over the multiplicative form (\gls{rmse} $11.20$, \gls{corr} $0.348$) and comparable to the static optimum GPMpostNP ($7.31$).

The three experiments converge on one conclusion: antagonism is primarily a structural consequence of multiplicative correction under demand conservation constraints, not of signal redundancy or double-normalization.
When the base allocation already captures the dominant spatial demand pattern, as the \gls{gnn} does, multiplicative factors systematically distort the distribution, amplifying high-demand areas while compressing low-demand areas within a zero-sum budget.

\paragraph{Embedding probing}
To test directly whether the \gls{gnn} internally encodes \gls{ntl} or Proximity information, which is the premise underlying the signal-redundancy hypothesis, $128$-dimensional agent-node embeddings are extracted from trained models ($48$ region--seed--fold combinations) and their correlation with the correction factors is computed (Tab.~\ref{tab:embedding_probing}).

The maximum per-dimension Pearson correlation between embeddings and \gls{ntl} is $|r| < 0.07$ on average; for Proximity, $|r| < 0.08$.
The \gls{gnn} encoder does not systematically encode either signal in its learned representations.
The final allocation weights $w_{sa}$ do show moderate Spearman correlation with \gls{ntl} ($\rho \approx 0.36$) and weak correlation with Proximity ($\rho \approx 0.23$), but this is a structural consequence of the task, since both \gls{ntl} and optimal demand allocation reflect urbanization intensity, rather than evidence of redundant learning within the encoder.

\begin{table*}[t]
\centering
\caption{Mechanism isolation: diagnostic experiments on the \gls{gnn} base allocation.
All experiments use the same $3 \times 4 = 12$ seed--fold evaluation protocol; values are seed-averaged over three initializations ($16$-region mean $\pm$ inter-region std).
``Mult.'' = multiplicative post-correction with re-normalization; ``Add.'' = additive post-correction with re-normalization; ``No-renorm'' = multiplicative without re-normalization; ``Random'' = spatially uncorrelated random noise with re-normalization.
$\Delta$\gls{rmse} is relative to the uncorrected \gls{gnn} baseline.}
\label{tab:mechanism_isolation}
\small
\begin{tabular}{llcccc}
\toprule
\textbf{Experiment} & \textbf{Correction form} & \textbf{\gls{rmse}} & \textbf{$\Delta$\gls{rmse}} & \textbf{\gls{mae}} & \textbf{\gls{corr}} \\
\midrule
\multicolumn{6}{l}{\textit{Controls}} \\
\gls{gnn} baseline (no correction)        & ---         & $9.27 \pm 2.61$  & ---                   & $6.67 \pm 1.76$ & $0.298 \pm 0.201$ \\
\gls{gnn} + \gls{ntl}$\times$Prox (Mult.)       & Mult.+renorm & $11.20 \pm 2.64$ & $+1.93$ ($+20.8\%$)  & $7.64 \pm 1.66$ & $0.348 \pm 0.160$ \\
\gls{gnn} + \gls{ntl} only (Mult.)              & Mult.+renorm & $9.45 \pm 2.55$  & $+0.19$ ($+2.0\%$)   & $6.58 \pm 1.64$ & $0.365 \pm 0.187$ \\
\gls{gnn} + Prox only (Mult.)             & Mult.+renorm & $9.42 \pm 2.33$  & $+0.15$ ($+1.6\%$)   & $6.59 \pm 1.52$ & $0.349 \pm 0.166$ \\
\midrule
\multicolumn{6}{l}{\textit{Exp~1: Removing re-normalization}} \\
No-renorm \gls{ntl}$\times$Prox           & Mult.        & $17.53 \pm 6.75$ & $+8.26$ ($+89.1\%$)  & $10.71 \pm 3.34$ & $0.357 \pm 0.167$ \\
No-renorm \gls{ntl} only                  & Mult.        & $10.37 \pm 2.82$ & $+1.10$ ($+11.9\%$)  & $7.10 \pm 1.87$ & $0.368 \pm 0.187$ \\
No-renorm Prox only                 & Mult.        & $11.62 \pm 4.20$ & $+2.35$ ($+25.4\%$)  & $7.63 \pm 2.22$ & $0.357 \pm 0.170$ \\
\midrule
\multicolumn{6}{l}{\textit{Exp~2: Random-noise post-correction}} \\
Random noise (10 repeats avg.)      & Mult.+renorm & $9.59 \pm 2.68$  & $+0.33$ ($+3.5\%$)   & $6.88 \pm 1.82$ & $0.287 \pm 0.199$ \\
\midrule
\multicolumn{6}{l}{\textit{Exp~3: Additive post-correction}} \\
Additive \gls{ntl}$\times$Prox            & Add.+renorm  & $\mathbf{7.34 \pm 2.14}$  & $\mathbf{-1.92}$ ($-20.8\%$)  & $5.37 \pm 1.41$ & $0.448 \pm 0.188$ \\
Additive \gls{ntl} only                   & Add.+renorm  & $8.82 \pm 2.63$  & $-0.45$ ($-4.8\%$)   & $6.35 \pm 1.84$ & $0.452 \pm 0.184$ \\
Additive Prox only                  & Add.+renorm  & $\mathbf{7.03 \pm 2.27}$  & $\mathbf{-2.24}$ ($-24.1\%$)  & $5.20 \pm 1.46$ & $0.405 \pm 0.164$ \\
\bottomrule
\end{tabular}
\end{table*}

\begin{table*}[t]
\centering
\caption{Embedding probing: correlation between \gls{gnn} internal representations and post-correction factors ($48$ region--seed--fold combinations per config).
``Embedding $r^{*}$'' reports the mean signed Pearson $r$ of the single embedding dimension attaining the largest $|r|$, selected over all $128$ dimensions and averaged across region--seed--fold combinations.
``Weight $\rho$'' reports the mean Spearman correlation between per-source allocation weights $w_{sa}$ and the factor.}
\label{tab:embedding_probing}
\small
\begin{tabular}{llcccc}
\toprule
 & & \multicolumn{2}{c}{\textbf{\gls{ntl} factor}} & \multicolumn{2}{c}{\textbf{Proximity factor}} \\
\cmidrule(lr){3-4} \cmidrule(lr){5-6}
\textbf{\gls{gnn} config} & \textbf{$n$} & Emb.\ $r^{*}$ & Weight $\rho$ & Emb.\ $r^{*}$ & Weight $\rho$ \\
\midrule
Baseline (no prior)  & 48 & $0.069 \pm 0.616$  & $0.360 \pm 0.083$ & $0.040 \pm 0.360$  & $0.226 \pm 0.080$ \\
\gls{ntl}+Prox prior       & 48 & $0.020 \pm 0.615$  & $0.390 \pm 0.121$ & $-0.078 \pm 0.353$ & $0.258 \pm 0.140$ \\
\bottomrule
\end{tabular}

\vspace{0.5em}
\footnotesize
\textit{Note}: $r^{*}$ is the \emph{signed} Pearson $r$ of the dimension selected by $\max|r|$; its mean is near zero because the sign of the most-correlated dimension varies across regions, while the standard deviation ($\approx 0.6$ for \gls{ntl}) reflects the typical magnitude of the strongest single-dimension correlation.
No single embedding dimension consistently encodes \gls{ntl} or Proximity across regions.
\end{table*}

\subsection{Sensitivity to Correction Strength}
\label{sec:sensitivity}

The \gls{gnn} is trained under weak supervision without substation-level demand labels.
In the deployment settings this method targets, such labels are generally unavailable,
so neither the post-correction parameters ($\alpha$, $\gamma$) nor the prior-loss weight ($\lambda$) can be tuned against disaggregation \gls{rmse}.
All hyperparameters in the main experiments were fixed \emph{a priori}:
$\alpha = 1$ and $\gamma = 2$ follow literature defaults;
$\lambda = 0.05$ balances the converged magnitudes of the loss terms
(land-use reconstruction $\approx 0.01$; each prior loss $\approx 0.7$--$0.8$).
The sweeps below do not seek optimal values; they verify that the pre-specified defaults do not fall in a pathological regime.

\begin{figure}[htbp]
\centering
\includegraphics[width=\columnwidth]{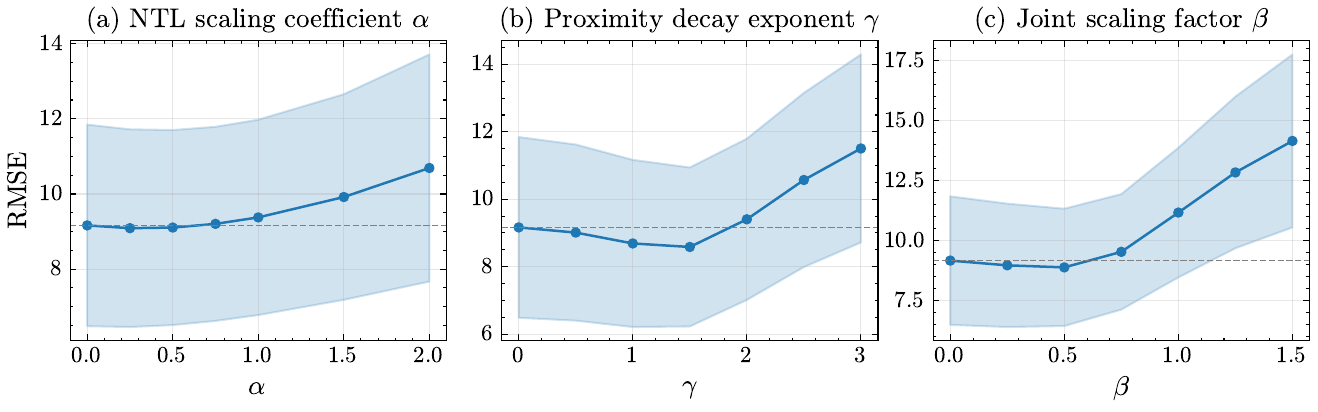}
\caption{Post-correction intensity sweep on the \gls{gnn} base.
(a)~\gls{ntl} scaling coefficient~$\alpha$.
(b)~Proximity decay exponent~$\gamma$.
(c)~Joint scaling factor~$\beta$ for the combined \gls{ntl}+Proximity correction.
The dashed line marks the uncorrected \gls{gnn} baseline.}
\label{fig:sweep}
\end{figure}

The \gls{ntl} scaling coefficient~$\alpha$, the Proximity decay exponent~$\gamma$, and a joint scaling factor~$\beta$ are each varied across seven levels; every sweep point is evaluated over all $3 \times 4 = 12$ seed--fold combinations (full results in~\ref{app:sweep}).
Single-factor sweeps show shallow, non-significant optima at low intensity
($\alpha = 0.25$: \gls{rmse}~$9.19$; $\gamma = 1.5$: \gls{rmse}~$8.65$; baseline $9.27$).
The combined sweep deteriorates steeply past $\beta = 0.5$, reaching $11.20$ at $\beta = 1.0$ ($+20.8\%$; Fig.~\ref{fig:sweep}).
Several correction intensities numerically improve RMSE on the GNN base, but none yield a statistically significant improvement.

A sweep over $\lambda \in \{0.01, 0.05, 0.1, 0.2, 0.5\}$ is conducted on a single fold (seed~42, fold~1, four regions) with retraining at each weight (Tab.~\ref{tab:sweep_b}).
Within the functional range $\lambda \in [0.01, 0.1]$, \gls{rmse} varies by less than one unit;
At $\lambda \geq 0.2$, the prior constraint dominates the main loss, and convergence degrades.
The default $\lambda = 0.05$ sits in the functional range and was not selected to minimize \gls{rmse}.
Both sweeps confirm that the antagonism documented in Sections~\ref{sec:asymmetry}--\ref{sec:mechanism_isolation} is not an artifact of pathological hyperparameter choices.

\subsection{Regional Heterogeneity}
\label{sec:heterogeneity}

To identify conditions under which the synergy--antagonism patterns vary, the $16$ regions are cross-stratified by load density ($\rho_r = D_r / \text{Area}_r$; terciles: low ${\leq}\,0.27$, mid $0.31$--$0.40$, high ${\geq}\,0.41$\,MVA\,km$^{-2}$) and land-use diversity (Shannon entropy $H_r$; median split at $H_r \approx 1.0$: low ${\leq}\,1.00$, high ${\geq}\,1.04$).
Tab.~\ref{tab:heterogeneity} reports the \gls{rmse} for six methods within each subgroup.
With $n = 16$ regions and a $3 \times 2$ cross-stratification, cell sizes range from 1 to 4 regions; the table should therefore be read as descriptive, flagging potential interaction patterns, rather than as definitive evidence of heterogeneity.

\begin{table*}[htbp]
\centering
\caption{\gls{rmse} by region subgroup (load density $\times$ land-use diversity).
The $n$ column gives the number of regions per cell; cells with $n = 1$ report a single region and carry no statistical weight.
GPMpostNP achieves the lowest \gls{rmse} in all six subgroups.}
\label{tab:heterogeneity}
\small
\begin{tabular}{llccccccc}
\toprule
\textbf{Density} & \textbf{Diversity} & $n$ & \textbf{Uni} & \textbf{GPMpostNP} & \textbf{\gls{gnn}} & \textbf{GNNpostP} & \textbf{GNNpostNP} & \textbf{GNNpriorNP} \\
\midrule
Low  & Low  & 4 & 12.56 & \textbf{7.88} & 9.29  & 10.41 & 13.02 & 9.32  \\
Low  & High & 1 & 18.30 & \textbf{6.90} & 11.94 & 11.40 & 13.02 & 11.68 \\
Mid  & Low  & 2 & 12.55 & \textbf{6.94} & 9.09  & 9.56  & 11.86 & 8.66  \\
Mid  & High & 4 & 11.19 & \textbf{6.56} & 7.90  & 8.23  & 9.99  & 7.55  \\
High & Low  & 2 & 13.07 & \textbf{5.77} & 8.16  & 8.14  & 9.87  & 7.44  \\
High & High & 3 & 15.02 & \textbf{8.96} & 11.04 & 9.77  & 10.24 & 11.04 \\
\bottomrule
\end{tabular}
\end{table*}

Four patterns emerge from this stratification.
Throughout this subsection, percentage gaps are computed as $(\text{RMSE}_A - \text{RMSE}_B) / \text{RMSE}_B$, where $B$ is the comparison reference (the method being compared against).

First, GPMpostNP ranks first in all six subgroups, with \gls{rmse} ranging from $5.77$ to $8.96$, whereas GNNpriorNP spans $7.44$ to $11.68$. The static pipeline owes its robustness to the fact that its correction factors are computed at the agent level, allowing them to adapt to local spatial structure without requiring the model to generalize across heterogeneous regions.

Second, the gap between \gls{gnn}-based and static methods depends strongly on region type.
In the most favourable setting for the \gls{gnn} (high density, low diversity; $n = 2$), GNNpriorNP trails GPMpostNP by $29\%$ (($7.44$ $-$ $5.77$)\,/\,5.77).
The low-density, low-diversity cell ($n = 4$) exhibits a comparable pattern: GNNpriorNP ($9.32$) versus GPMpostNP ($7.88$), an $18\%$ gap.
The low-density, high-diversity cell ($n = 1$, region TLC1 only) records the widest gap at $69\%$ (($11.68$ $-$ $6.90$)\,/\,6.90), however as a single data point it is of limited significance in isolation and merely consistent with the $n = 4$ cell.
Low-density regions in general present complex land-use mosaics with small absolute loads, conditions under which the \gls{gnn}'s five-dimensional feature space lacks the granularity to distinguish fine-grained demand patterns, whereas the Proximity factor directly encodes demand-hotspot information derived from substation siting.

Third, the high-density, high-diversity subgroup ($n = 3$: London, TLE4, TLD3) is anomalous.
GPMpostNP records its worst \gls{rmse} here ($8.96$), and GNNpriorNP ($11.04$) offers no improvement over the uncorrected \gls{gnn} ($11.04$).
This subgroup combines concentrated load with high spatial heterogeneity, conditions under which \gls{ntl} and Proximity signals may conflict; the pattern is consistent with \gls{ntl} indicating uniformly high activity and Proximity scores being diluted by the dense substation network, though the cell size ($n = 3$) precludes a definitive assessment.
London dominates the cell average (\gls{gnn} \gls{rmse} $= 16.9$, versus ${\sim}8$ for TLE4 and TLD3).
Excluding London, GNNpriorNP drops to $7.56$ (versus \gls{gnn} $8.09$), which restores the prior-loss advantage observed in other subgroups; the apparent anomaly is therefore attributable to London rather than to a structural property of the high-density, high-diversity condition.

Fourth, the combined post-correction antagonism persists across all subgroups.
GNNpostNP approaches or exceeds the Uniform baseline \gls{rmse} in every cell, with the most severe cases in low-density regions.
This rules out the hypothesis that antagonism is confined to particular region types.

\subsection{Cross-Network Transferability: German B\"{o}rde Case}
\label{sec:germany}

The German B\"{o}rde district ($13$ substations, $34$ Gemeinden) serves as a qualitative transferability check.
Given the small sample size, no statistical inference is attempted; the analysis focuses on whether the directional patterns observed in Britain replicate under a different network topology.
A jackknife leave-one-out check confirms that baseline GNN performance is not an artifact of overfitting or of a single dominant node: removing any one substation leaves LOO Corr~$\geq$~0.941, with the largest drop at Zielitz (45\,MW; minimum LOO $r = 0.919$).

\begin{table}[htbp]
\centering
\caption{Method performance in the German B\"{o}rde district (single-sample evaluation, three-seed mean $\pm$ std for \gls{gnn} methods).}
\label{tab:germany}
\begin{tabular}{lcc}
\toprule
\textbf{Method}  & \textbf{\gls{rmse}} & \textbf{\gls{corr}} \\
\midrule
Uni              & 6.50          & 0.860 \\
\gls{gpm}              & 6.88          & 0.846 \\
UniNP            & 1.95          & 0.984 \\
GPMpostNP        & 2.33          & 0.977 \\
\midrule
\gls{gnn}              & 2.49 $\pm$ 0.38 & 0.976 $\pm$ 0.007 \\
GNNpriorNP       & 2.77 $\pm$ 0.10 & 0.970 $\pm$ 0.002 \\
\bottomrule
\end{tabular}
\end{table}

Two of the three core British findings replicate qualitatively.

\paragraph{Static synergy confirmed}
UniNP ($1.95$) records a $22\%$ lower \gls{rmse} than \gls{gnn} ($2.49$) (($2.49 - 1.95$)\,/\,$2.49$), consistent with the $19\%$ gap in Britain (GPMpostNP versus GNNpriorNP).
The synergy of \gls{ntl} and Proximity on static bases is therefore not contingent on a particular grid topology.

\paragraph{GPM ranking does not replicate}
In Britain, GPMpostNP ($7.31$\,MVA) records a lower \gls{rmse} than UniNP ($7.45$\,MVA); in B\"{o}rde the ranking is reversed, with UniNP ($1.95$\,MVA) recording a lower \gls{rmse} than GPMpostNP ($2.33$\,MVA).
This reversal is consistent with the single-sample nature of the German evaluation and with the coarse spatial resolution of the economic output data that supplies the \gls{gpm} weighting signal, so whether the ranking reversal reflects a systematic limitation of land-use weighting or a case-specific artifact cannot be assessed here.

\paragraph{Post-correction antagonism amplified}
\gls{ntl} post-correction worsens the German \gls{gnn} baseline from $2.49$ to $2.99$ ($+20\%$), compared with $+2\%$ in Britain; the combined \gls{ntl}+Proximity post-correction inflates \gls{rmse} to $5.98$ ($+140\%$), compared with $+21\%$ in Britain.
In a small, spatially homogeneous network where the \gls{gnn} already captures nearly all learnable variance, multiplicative perturbations translate more directly into error because there is less residual structure for the correction to exploit.

A jackknife leave-one-out analysis across methods shows that multiplicative correction degrades robustness to single-point removal alongside average accuracy.
The standard deviation of LOO correlation rises from 0.013 (uncorrected GNN) through 0.025 and 0.033 (single post-corrections) to 0.048 (combined NTL+Proximity); the minimum LOO correlation falls from 0.919 to 0.710.
Removing one substation from the combined post-correction can reduce correlation by 0.17 points, a level of instability incompatible with operational settings where individual metering outages are routine.

\paragraph{Prior-loss boundary condition exposed}
Unlike in Britain, dual prior-loss training worsens the German \gls{gnn}: \gls{rmse} rises from $2.49$ to $2.77$ ($+11.2\%$).
The per-component breakdown (Tab.~\ref{tab:prior_boundary}) shows that all three prior configurations degrade performance.

\begin{table}[htbp]
\centering
\caption{Prior-loss effect comparison between Britain and Germany.
The sign reversal in Germany exposes the boundary condition: when the \gls{gnn} baseline already explains more than $95\%$ of demand variance ($r^2 = 0.953$), prior constraints impose a net optimization burden.}
\label{tab:prior_boundary}
\begin{tabular}{lcc}
\toprule
\textbf{Prior config.} & \textbf{Britain $\Delta$\gls{rmse}} & \textbf{Germany $\Delta$\gls{rmse}} \\
\midrule
\gls{ntl} only      & $-0.08$ ($-0.9\%$)  & $+0.13$ ($+5.1\%$)   \\
Prox only     & $-0.13$ ($-1.4\%$)  & $+0.06$ ($+2.6\%$)   \\
\gls{ntl}+Prox      & $-0.24$ ($-2.6\%$)  & $+0.28$ ($+11.2\%$)  \\
\bottomrule
\end{tabular}
\end{table}

The divergence is readily explained by the difference in baseline performance levels.
The British \gls{gnn} baseline operates at $r = 0.298$, leaving substantial room for the prior to supply useful inductive bias.
The German \gls{gnn} baseline sits at $r = 0.976$, leaving almost no residual structure to exploit.
At this operating point, the additional loss terms introduce gradient competition that outweighs their informational benefit.
This establishes a necessary condition for prior-loss utility: the residual prediction gap ($1 - r^2$, the unexplained variance fraction) must be large enough for the prior to contribute information that the main loss alone does not capture.
Concretely, the British baseline ($r = 0.298$, $1 - r^2 = 0.911$) leaves $91\%$ of demand variance unexplained, whereas the German baseline ($r = 0.976$, $1 - r^2 = 0.047$) leaves only $5\%$.
The condition is necessary but not sufficient: the British case satisfies it, yet the prior-loss improvement remains non-significant ($p = 0.67$), indicating that a large residual gap is required but, by itself, does not guarantee benefit.

The \gls{ntl}--Proximity correlation in the German case is $\rho = 0.258$ ($p = 0.556$), but with $N = 13$ substations, the test has negligible statistical power; the redundancy-amplification hypothesis cannot be confirmed or refuted here.

\paragraph{Downstream system impact}
Tab.~\ref{tab:downstream} summarizes the power-flow outcomes when each method's demand estimates are injected into the simplified 110\,kV network (Section~\ref{subsec:germany_setup}). 
$\Delta_{\text{MAE}}$ broadly scales with RMSE across the method spectrum, spanning a 23-fold range from UniNP ($0.74$\,pp) to Flat ($17.03$\,pp). 
The relationship is not strictly monotonic, but the overall trend holds across the range of methods tested. 
In this radial topology with a single geographically remote high-demand node, spatial demand conservation induces an ordering inversion: absolute maximum line loading decreases with higher RMSE ($132.1\%$ for UniNP versus $102.5\%$ for the flat baseline), so $\Delta_{\text{MAE}}$ is the appropriate method-comparison metric. 
This inversion is specific to radial networks with a dominant distal node and does not generalize to meshed topologies. 

\begin{table}[htbp]
\centering
\caption{Downstream power-flow outcomes for the B\"{o}rde simplified 110\,kV network (methods sorted by RMSE ascending).
$\Delta_{\text{MAE}}$: per-line mean absolute deviation of line loading from the
true-load reference (pp).
$\ell_{\max}$: absolute maximum line loading (\%).
Flat: each substation receives an equal share of the total district demand.}
\label{tab:downstream}
\begin{tabular}{lcccc}
\toprule
\textbf{Method} & \textbf{RMSE (MVA)} & $\boldsymbol{\Delta_{\text{MAE}}}$\,\textbf{(pp)} & $\boldsymbol{\ell_{\max}}$\,\textbf{(\%)} \\
\midrule
True load             & $---$   & $0.00$  & $132.9$ \\
UniNP                 & $1.95$  & $0.74$  & $132.1$\\
\gls{gnn}             & $2.49$  & $1.39$  & $129.3$ \\
GNNpriorNP            & $2.77$  & $1.57$  & $129.4$ \\
Flat                  & $10.77$ & $17.03$ & $102.5$ \\
\bottomrule
\end{tabular}
\end{table}
\section{Discussion}
\label{sec:discussion}

The results present an apparent paradox: a \gls{gnn} trained end-to-end outperforms hand-crafted \gls{gpm} weighting by a significant margin, yet the best overall method is a static pipeline that layers \gls{gpm}, \gls{ntl}, and Proximity corrections without any learned component.
This section interprets the three central phenomena (mechanism-dependent antagonism, \gls{rmse}--\gls{corr} decoupling, and prior-loss boundary conditions) and provides practical guidance.

\subsection{Mechanistic Interpretation}

\paragraph{Structural origin of post-correction antagonism}
The divergent behavior documented in Sections~\ref{sec:asymmetry}--\ref{sec:mechanism_isolation} is not a matter of degree but of mathematical structure.
A static base encodes almost no spatial demand signal; every multiplicative correction factor therefore injects new information into a near-uniform distribution, and the conservation constraint merely rescales the improved allocation.
A \gls{gnn} base, by contrast, has already concentrated probability mass on a subset of substations.
Under the same conservation constraint, multiplicative rescaling converts every local amplification into compensating compression elsewhere, systematically distorting the allocation the network learned to produce.
The critical insight is that the correction form, not the correction content, determines whether auxiliary information helps or harms.
Multiplicative post-correction is the de facto standard in spatial disaggregation, inherited from an era when base allocations were rule-based; whenever a learning-based component replaces a deterministic one, the mathematical assumptions embedded in downstream post-correction steps require re-examination.

When corrections are applied sequentially, this distortion amplifies geometrically: the combined \gls{ntl}+Proximity correction degrades \gls{gnn} output far more than the sum of individual degradations ($+1.93$ versus $+0.34$), because each multiplicative factor compounds the distribution distortion introduced by its predecessor under the fixed-sum constraint.
The \gls{ntl}--Proximity correlation ($\rho = 0.48$) amplifies this damage on a learned base but does not cause it; on a static base, the same correlation produces near-perfect additivity.

\paragraph{Prior-loss integration: non-harmful but bounded}
Prior-loss training sidesteps antagonism by embedding auxiliary information into the optimization objective, allowing gradient-based adjudication of conflicts with the main loss.
Two structural constraints bound its benefit: the proxy-target alignment gap and residual baseline capacity.
The proxy-target alignment gap arises because the self-supervised land-use reconstruction loss need not minimize demand \gls{rmse}; this misalignment accounts for the persistent shortfall against the static optimum and implies that closing it, rather than adding more auxiliary losses, is the productive direction.
The residual baseline capacity constraint emerges when baseline performance is near ceiling: additional loss terms then introduce gradient competition rather than useful inductive bias, as the German case illustrates.
The necessary condition for prior-loss utility is that $1 - r^2$ be large enough for the prior to supply structure that the main loss alone does not capture; this is necessary but not sufficient.

\paragraph{\gls{rmse}--\gls{corr} decoupling and its implications for evaluation}
The observation that post-correction can improve discrimination while worsening calibration is consistent with three jointly sufficient conditions: a conservation constraint, multiplicative correction, and a base allocation that already captures the dominant spatial pattern.
Under these conditions, multiplicative rescaling stretches the demand distribution, improving rank ordering while inflating absolute errors at the tails.
This is not a pathology specific to energy disaggregation; any constrained spatial allocation problem where corrections are applied multiplicatively to a pre-trained model is susceptible.
Reporting only Pearson or Spearman correlation would mask the calibration penalty observed here and could lead to the adoption of correction schemes that worsen actual allocation accuracy.

\subsection{Engineering Significance and Practical Guidance}

\paragraph{Planning margin reduction}
The improvement from the Uniform baseline (Uni, $13.10$~MVA) to the static optimum (GPMpostNP, $7.31$~MVA) corresponds to a $44\%$ reduction (($13.10 - 7.31$)\,/\,$13.10$) in average substation-level demand estimation error.
Typical primary substation transformers in Great Britain are rated at $30$--$90$~MVA; reducing the estimation uncertainty from $13$~MVA to $7$~MVA narrows the required planning margin by approximately $6$~MVA per substation, enabling tighter capacity allocation and potential deferral of reinforcement investment.
The simplified 110\,kV power-flow test on the B\"{o}rde district (Tab.~12) shows that disaggregation errors amplify through downstream calculations: the RMSE ratio between UniNP and the flat baseline is 5.5:1, but the corresponding ratio in mean line-loading deviation is 23:1 (0.74 versus 17.03 percentage points).
Inaccurate methods also underestimate the maximum line loading in this radial topology (102.5\,\% versus the true 132.9\,\%), because spreading demand away from the dominant distal node removes the overload signal that would trigger reinforcement.
Multiplicative post-correction compounds the problem: the jackknife analysis shows that removing a single substation can drop the leave-one-out correlation from 0.919 to 0.710, which rules out this correction form in any setting where metering outages occur.
 
\paragraph{Deployment guidance}
Three practical recommendations follow.
First, without substation-level ground truth for \gls{gnn} training, which is the norm in most planning contexts, use the static three-layer pipeline (Voronoi partition, \gls{gpm} weighting, Proximity and \gls{ntl} correction).
It provides the strongest disaggregation at zero training cost, using freely available data (VIIRS DNB, OpenStreetMap).
Second, when a \gls{gnn} is deployed, avoid multiplicative post-correction; additive post-correction is a candidate alternative that eliminates antagonism in the present experiments (Section~\ref{sec:mechanism_isolation}), though the optimal correction form remains to be established across broader settings.
Prior-loss integration is a second option where residual model capacity exists, but should be validated by ablation before deployment.
Third, diagnose prior-loss viability from the baseline's $1 - r^2$: the German case shows that when this quantity is small, additional loss terms introduce gradient competition rather than useful inductive bias.

\paragraph{Implications for open-source energy system models}
\gls{ntl} captures fine-grained spatial variation in economic activity below the NUTS 3 resolution at which GDP is typically reported, while Proximity encodes network-topology information absent from demographic data.
Both layers, therefore, inject a non-redundant signal into any Voronoi-based pipeline that currently relies solely on socioeconomic weighting, at negligible computational cost and with no model retraining required, making them directly applicable to existing open-source frameworks such as PyPSA-Eur~\cite{Horsch2018pypsa}.

\subsection{Limitations}
\label{sec:limitations}

\paragraph{Geographic scope}
The primary analysis covers Great Britain ($16$ \gls{itl}-2 regions), supplemented by one small German district ($13$ substations).
Both networks share Western European characteristics: moderate load density, high data availability, and temperate climate.
Networks with fundamentally different properties, such as dispersed rural grids in sub-Saharan Africa, high-density monsoon-climate systems in Southeast Asia, or generation-dominated topologies where substation siting does not follow load centers, would stress different parts of the pipeline and could invert the relative ranking of methods.
The German case, while qualitatively confirmatory, is too small ($n = 13$) for statistical inference and cannot substitute for full-scale cross-network replication.

\paragraph{\gls{gnn} architecture and training regime}
The \gls{gnn} architecture (\gls{hgt}, five land-use features, self-supervised land-use reconstruction) is fixed throughout.
This is both a strength, as it isolates the effect of auxiliary information integration from confounding architectural decisions, and a limitation.
Richer input features (building footprints, fine-resolution socioeconomic indicators), alternative graph convolution operators, or supervised training on demand labels could shift the \gls{gnn} baseline closer to the static optimum, narrowing or eliminating the gap.
Critically, the proxy-target alignment gap identified above is specific to the self-supervised objective; a supervised \gls{gnn} would bypass it entirely, potentially changing the prior-loss boundary condition.

\paragraph{Auxiliary data quality}
The static pipeline's advantage partly reflects the quality of the available \gls{ntl} (VIIRS DNB) and Proximity data for Great Britain.
Two failure modes are foreseeable: in low-electrification regions, \gls{ntl} radiance saturates near zero, collapsing its discriminative power; in networks where substations serve generation rather than load, the Proximity factor encodes the wrong spatial signal.
The moderate \gls{ntl}--Proximity correlation ($\rho = 0.48$) observed in Britain is itself context-dependent and would alter the super-additive antagonism pattern if substantially higher or lower.

\paragraph{Temporal scope}
All demand values represent annual peak load.
Temporal disaggregation, which distributes load profiles across hours, days, or seasons, introduces dynamics orthogonal to the spatial structure examined here.
The conservation constraint that drives the antagonism mechanism operates identically at each time step, so the structural finding is expected to transfer; whether its practical magnitude remains significant under temporal variation is untested.

\paragraph{Correction functional form}
Only multiplicative and additive post-correction are examined.
Residual correction ($\hat{d}_j \leftarrow \hat{d}_j + f_\theta(\mathbf{x}_j)$), attention-weighted adjustment, or learned post-processors could offer different trade-offs between antagonism avoidance and information utilization.
The additive result establishes that the antagonism is form-dependent, but does not identify the optimal functional form; it shows only that the standard multiplicative form is the wrong one for learned bases.

\section{Conclusion}
\label{sec:conclusion}

This paper tested whether auxiliary spatial information that improves rule-based demand weighting still helps once the weighting component is replaced by a learned model.
We fixed Voronoi partitioning and systematically crossed two design axes, namely demand weighting (static Grid Point Model versus graph neural network) and auxiliary-information integration (multiplicative post-correction applied after inference versus prior-loss terms injected during training), across 15 configurations evaluated on metered peak demand from 1{,}891 British primary substations, with a directional transfer check on 13 German substations.
 
The answer depends on the integration mechanism.
Multiplicative post-correction with Nighttime Light intensity and substation-proximity data reduces root-mean-square error by $41\,\%$ on the static base but increases it by $21\,\%$ on the graph neural network base ($p < 0.001$ for both), leaving the best static variant $19\,\%$ below the best learned variant in absolute error.
The cause is structural: under demand conservation, multiplicative rescaling distorts an already informative learned distribution by converting local amplifications into compensating compressions elsewhere.
Post-correction also decouples calibration from discrimination, because rank-order correlation improves significantly ($p < 0.001$) while absolute error worsens, so correlation-only evaluation masks the penalty.
 
Prior-loss training avoids the antagonism but yields only a non-significant improvement ($-2.6\,\%$) on the British data; in the German case, where the learned baseline already explains over $95\,\%$ of demand variance, prior-loss training shows signs of degradation, exposing a boundary condition tied to residual model capacity.
 
The Nighttime Light and substation-proximity signals behind the $41\,\%$ static improvement are publicly available, yet most open-source energy system frameworks rely solely on socioeconomic weighting.
Without substation-level ground truth for neural-network training, which is the norm in most planning contexts, the three-layer static pipeline (Voronoi partition, land-use-based weighting, Nighttime Light and proximity correction) delivers the strongest disaggregation at zero training cost and should serve as the default.
When a learned weighting model is deployed, multiplicative post-correction should not carry over; additive correction eliminates the antagonism in the present experiments, though the optimal non-multiplicative form remains unestablished, and prior-loss integration offers an alternative where residual model capacity has not been exhausted.
Method evaluation should report root-mean-square error and correlation jointly, as the two metrics diverge under post-correction on learned representations.
 
The broader point is that method rankings in spatial disaggregation are not intrinsic to models but depend on how auxiliary information enters the allocation pipeline, a dependence that single-axis comparisons without ground-truth data cannot detect, yet the affected Voronoi-based allocation module is a widely reused upstream component in open-source energy system modeling.
Developing principled non-multiplicative correction frameworks and validating the antagonism pattern on networks with fundamentally different properties, such as dispersed rural grids, generation-dominated topologies, and tropical-climate demand structures, are the direct next steps.

\section*{Declaration}
During the preparation of this work, the authors used Gemini (Google DeepMind) and Claude (Anthropic) in order to perform text polishing and code debugging for plotting. After using this tool/service, the authors reviewed and edited the content as needed and assume full responsibility for the content of the publication.

\section*{Acknowledgments}

This work is supported in part by the Helmholtz Association through the project “Helmholtz Platform for the Design of Robust Energy Systems and Their Supply Chains” (RESUR) and the Helmholtz AI HAICORE partition (HAICORE@KIT).

\bibliographystyle{elsarticle-num-names}
\bibliography{cas-refs}

@article{Aryanpur2021review,
  author    = {Aryanpur, Vahid and {\'O}'Gallach{\'o}ir, Brian and Dai, Hancheng and Chen, Wenying and Glynn, James},
  title     = {A review of spatial resolution and regionalisation in national-scale energy systems optimisation models},
  journal   = {Energy Strategy Reviews},
  volume    = {37},
  pages     = {100702},
  year      = {2021},
  doi       = {10.1016/j.esr.2021.100702},
}

@article{Frysztacki2021strong,
  author    = {Frysztacki, Martha Maria and Horsch, Jonas and Hagenmeyer, Veit and Brown, Tom},
  title     = {The strong effect of network resolution on electricity system models with high shares of wind and solar},
  journal   = {Applied Energy},
  year      = {2021},
  note      = {Online first},
  doi       = {10.1016/J.APENERGY.2021.116726},
}

@article{Hulk2017allocation,
  author    = {H{\"u}lk, Ludwig and Wienholt, Lukas and Cu{\ss}mann, Ilka and M{\"u}ller, Ulf Philipp and Matke, Carsten and K{\"o}tter, Editha},
  title     = {Allocation of Annual Electricity Consumption and Power Generation Capacities across Multiple Voltage Levels in a High Spatial Resolution},
  journal   = {International Journal of Sustainable Energy Planning and Management},
  volume    = {13},
  pages     = {79--92},
  year      = {2017},
  month     = sep,
  doi       = {10.5278/ijsepm.2017.13.6},
}

@incollection{Koppl2017modeling,
  author    = {K{\"o}ppl, Simon and B{\"o}ing, Felix and Pellinger, Christoph},
  title     = {Modeling of the Transmission Grid Using Geo Allocation and Generalized Processes},
  booktitle = {Advances in Energy System Optimization},
  editor    = {Bertsch, Valentin and Fichtner, Wolf and Heuveline, Vincent and Leibfried, Thomas},
  publisher = {Springer International Publishing},
  year      = {2017},
  doi       = {10.1007/978-3-319-51795-7_12},
}

@article{JalilVega2018effect,
  author    = {Jalil-Vega, Francisca and Hawkes, Adam},
  title     = {The effect of spatial resolution on outcomes from energy systems modelling of heat decarbonisation},
  journal   = {Energy},
  volume    = {155},
  pages     = {339--350},
  year      = {2018},
  doi       = {10.1016/j.energy.2018.04.160},
}

@article{MartinezGordon2021review,
  author    = {Mart{\'i}nez-Gord{\'o}n, R. and Morales-Espa{\~n}a, G. and Sijm, J. and Faaij, A.},
  title     = {A review of the role of spatial resolution in energy systems modelling: Lessons learned and applicability to the {North Sea} region},
  journal   = {Renewable \& Sustainable Energy Reviews},
  volume    = {141},
  pages     = {110857},
  year      = {2021},
  doi       = {10.1016/j.rser.2021.110857},
}

@inproceedings{Mei2016spatial,
  author    = {Mei, Jiali and Goude, Yannig and Hebrail, Georges and Kong, Nicolas},
  title     = {Spatial estimation of electricity consumption using socio-demographic information},
  booktitle = {2016 IEEE PES Asia-Pacific Power and Energy Engineering Conference (APPEEC)},
  pages     = {753--757},
  year      = {2016},
  month     = {10},
  doi       = {10.1109/APPEEC.2016.7779596},
}

@article{Singh2015high,
  author    = {Singh, Antriksh and Eser, Patrick and Chokani, Ndaona and Abhari, Reza},
  title     = {High Resolution Modeling of the Impacts of Exogenous Factors on Power Systems---Case Study of {Germany}},
  journal   = {Energies},
  volume    = {8},
  number    = {12},
  year      = {2015},
  doi       = {10.3390/en81212424},
}

@article{Chen2022global,
  author    = {Chen, Jiandong and Gao, Ming and Cheng, Shulei and others},
  title     = {Global 1 km $\times$ 1 km Gridded Revised Real Gross Domestic Product and Electricity Consumption during 1992--2019 Based on Calibrated Nighttime Light Data},
  journal   = {Scientific Data},
  volume    = {9},
  number    = {1},
  pages     = {202},
  year      = {2022},
  doi       = {10.1038/s41597-022-01322-5},
}

@article{Crippa2024insights,
  author    = {Crippa, Monica and Guizzardi, Diego and Pagani, Federico and others},
  title     = {Insights into the Spatial Distribution of Global, National, and Subnational Greenhouse Gas Emissions in the {Emissions Database for Global Atmospheric Research} ({EDGAR} v8.0)},
  journal   = {Earth System Science Data},
  volume    = {16},
  number    = {6},
  pages     = {2811--2830},
  year      = {2024},
  doi       = {10.5194/essd-16-2811-2024},
}

@article{Danylo2019high,
  author    = {Danylo, Olha and Bun, Rostyslav and See, Linda and Charkovska, Nadiia},
  title     = {High-Resolution Spatial Distribution of Greenhouse Gas Emissions in the Residential Sector},
  journal   = {Mitigation and Adaptation Strategies for Global Change},
  volume    = {24},
  number    = {6},
  pages     = {941--967},
  year      = {2019},
  doi       = {10.1007/s11027-019-9846-z},
}

@unpublished{Patil2025spatially,
  author    = {Patil, Shruthi and Pflugradt, Noah and Weinand, Jann and Kropp, J{\"u}rgen and Stolten, Detlef},
  title     = {Spatially Disaggregated Energy Consumption and Emissions in End-Use Sectors for {Germany} and {Spain}},
  year      = {2025},
  note      = {arXiv preprint},
  doi       = {10.48550/arXiv.2505.05139},
}

@article{Patil2024systematic,
  author    = {Patil, Shruthi and Pflugradt, Noah and Weinand, Jann M. and Stolten, Detlef and Kropp, J{\"u}rgen},
  title     = {A systematic review of spatial disaggregation methods for climate action planning},
  journal   = {Energy and AI},
  volume    = {17},
  pages     = {100386},
  year      = {2024},
  month     = sep,
  doi       = {10.1016/j.egyai.2024.100386},
}

@article{Raventos2022comparison,
  author    = {Ravent{\'o}s, Oriol and Dengiz, Thomas and Medjroubi, Wided and Unaichi, Chinonso and Bruckmeier, Andreas and Finck, Rafael},
  title     = {Comparison of Different Methods of Spatial Disaggregation of Electricity Generation and Consumption Time Series},
  journal   = {Renewable and Sustainable Energy Reviews},
  volume    = {163},
  pages     = {112186},
  year      = {2022},
  month     = jul,
  doi       = {10.1016/j.rser.2022.112186},
}

@ARTICLE{Gong2019acomparison,
  author={Gong, Ting and Lee, Tyler and Stephenson, Cory and Renduchintala, Venkata and Padhy, Suchismita and Ndirango, Anthony and Keskin, Gokce and Elibol, Oguz H.},
  journal={IEEE Access}, 
  title={A Comparison of Loss Weighting Strategies for Multi task Learning in Deep Neural Networks}, 
  year={2019},
  volume={7},
  number={},
  pages={141627-141632},
  doi={10.1109/ACCESS.2019.2943604}}

@article{Bhattarai2023remote,
author = {Bhattarai, Dipendra and Lucieer, Arko and Lovell, Heather and Aryal, Jagannath},
title = {Remote sensing of night-time lights and electricity consumption: A systematic literature review and meta-analysis},
journal = {Geography Compass},
volume = {17},
number = {4},
pages = {e12684},
doi = {https://doi.org/10.1111/gec3.12684},
year = {2023}
}

@INPROCEEDINGS{Baran2005load,
  author={Baran, M.},
  booktitle={IEEE Power Engineering Society General Meeting, 2005}, 
  title={Load estimation for load monitoring at distribution substations}, 
  year={2005},
  volume={},
  number={},
  pages={902 Vol. 1-},
  doi={10.1109/PES.2005.1489653}}

@article{Borges2020enhancing,
title = {Enhancing the missing data imputation of primary substation load demand records},
journal = {Sustainable Energy, Grids and Networks},
volume = {23},
pages = {100369},
year = {2020},
issn = {2352-4677},
doi = {https://doi.org/10.1016/j.segan.2020.100369},
author = {Cruz E. Borges and Oihane Kamara-Esteban and Tony Castillo-Calzadilla and Cristina Martin Andonegui and Ainhoa Alonso-Vicario},
}

@inproceedings{Mu2026Improving,
      author={Mu, Xuanhao and Geiges, Jakob and Liu, Nan and Schlachter, Thorsten and Hagenmeyer, Veit},
      booktitle={14th Power Systems Computation Conference PSCC 2026}, 
      title={Improving Spatial Allocation for Energy System Coupling with Graph Neural Networks}, 
      month={Juni.},
      year={2026},
      note={in press},
      url={https://arxiv.org/abs/2602.22249}
}

@article{Horsch2018pypsa,
title = {PyPSA-Eur: An open optimisation model of the European transmission system},
journal = {Energy Strategy Reviews},
volume = {22},
pages = {207-215},
year = {2018},
issn = {2211-467X},
doi = {https://doi.org/10.1016/j.esr.2018.08.012},
author = {Jonas Hörsch and Fabian Hofmann and David Schlachtberger and Tom Brown},
}

@article{Chen2024Physics-Informed,
title={Physics-Informed Neural Networks with Hard Linear Equality Constraints},
author={Hao Chen and Gonzalo E. Constante-Flores and Canzhou Li},
journal={Comput. Chem. Eng.},
year={2024},
volume={189},
pages={108764},
doi={10.48550/arxiv.2402.07251}
}

@article{Liang2022Knowledge,
title={Knowledge Graph Contrastive Learning Based on Relation-Symmetrical Structure},
author={K. Liang and Yue Liu and Sihang Zhou and Wenxuan Tu and Yi Wen and Xihong Yang and Xiang Dong and Xinwang Liu},
journal={IEEE Transactions on Knowledge and Data Engineering},
year={2022},
volume={36},
pages={226-238},
doi={10.1109/tkde.2023.3282989}
}

@INPROCEEDINGS{Mu2025Improving,
  author={Mu, Xuanhao and Geiges, Jakob and Liu, Jianlei and Schlachter, Thorsten and Hagenmeyer, Veit},
  booktitle={2025 IEEE 13th International Conference on Smart Energy Grid Engineering (SEGE)}, 
  title={Improving Spatial Allocation for Energy System Coupling with Clustering-Based Voronoi Diagrams}, 
  year={2025},
  volume={},
  number={},
  pages={195-200},
  doi={10.1109/SEGE65970.2025.11203477}}

@book{crato_figuring_2010,
	address = {Berlin, Heidelberg},
	title = {Figuring {It} {Out}},
	copyright = {http://www.springer.com/tdm},
	isbn = {978-3-642-04832-6 978-3-642-04833-3 },
	language = {en},
	urldate = {2024-10-24},
	publisher = {Springer},
	author = {Crato, Nuno},
	year = {2010},
	doi = {10.1007/978-3-642-04833-3},
}

@article{Zhou2024Dataset,
title = {Datasets of Great Britain primary substations integrated with household heating information},
journal = {Data in Brief},
volume = {54},
pages = {110483},
year = {2024},
issn = {2352-3409},
doi = {10.1016/j.dib.2024.110483},
author = {Yihong Zhou and Chaimaa Essayeh and Thomas Morstyn},
}

@misc{uk_gva,
title = {Regional gross value added (balanced) by industry: all ITL regions},
author = {{Office for National Statistics}},
year = {2024},
url = {https://www.ons.gov.uk/economy/grossvalueaddedgva},
note = {Accessed: 2025-04-04}
}

@misc{uk_population,
title = {Estimates of the population for England and Wales},
author = {{Office for National Statistics}},
year = {2024},
url = {https://www.ons.gov.uk/peoplepopulationandcommunity/populationandmigration/populationestimates},
note = {Accessed: 2025-04-04}
}

@misc{uk_itl,
title = {International Territorial Level},
author = {{Office for National Statistics}},
year = {2024},
url = {https://www.ons.gov.uk/methodology/geography/ukgeographies/eurostat},
note = {Accessed: 2025-08-27}
}

@misc{openstreetmap_contributors_planet_2017,
	title = {Planet dump retrieved from https://planet.osm.org},
	author = {{OpenStreetMap contributors}},
	year = {2017},
        url = {https://www.openstreetmap.org}
}

@misc{regiocom2026avacon,
    author       = {{regiocom SE}},
    title        = {{GeoServer Spatial Data Service --- Landkreis Börde Electrical Grid Infrastructure}}, 
    year         = {2024},
    howpublished = {Internal GeoServer WFS, \url{https://webapp.regiocom.net}}, 
    note         = {Layers used: \texttt{v\_epa\_uw} (substation locations),
                    \texttt{v\_epa\_uw\_last} (substation peak loads),
                    \texttt{v\_lkb\_vg250\_gem\_sql} (Gemeinde boundaries),
                    \texttt{v\_lkb\_vg250\_krs\_sql} (Kreis boundary).
                    Accessed Mar 2026},
    }

@misc{elvidge2017viirs,
    author    = {{Earth Observation Group (EOG), NOAA/NCEI}},
    title     = {{VIIRS Day/Night Band Nighttime Lights Monthly Composites, Version 1}},
    year      = {2017--present},
    note      = {Collection ID: \texttt{NOAA/VIIRS/DNB/MONTHLY\_V1/VCMSLCFG}.
                 Accessed via Google Earth Engine, median composite Jan 2021--Dec 2022},
    url       = {https://developers.google.com/earth-engine/datasets/catalog/NOAA_VIIRS_DNB_MONTHLY_V1_VCMSLCFG},
  }

@misc{eurostat_nama10r3gva,
  author       = {{Eurostat}},
  title        = {Gross Value Added at Basic Prices by {NUTS}~3 Regions},
  year         = {2026},
  howpublished = {\url{https://ec.europa.eu/eurostat/databrowser/view/nama_10r_3gva/default/table?lang=en}},
  note         = {Online data code: \texttt{nama\_10r\_3gva}. Accessed: 2026-03-13}
}

@misc{eurostat_gisco_nuts,
  author       = {{Eurostat -- GISCO}},
  title        = {{NUTS} -- Territorial Units for Statistics: Boundary Dataset},
  year         = {2024},
  howpublished = {\url{https://ec.europa.eu/eurostat/web/gisco/geodata/statistical-units/territorial-units-statistics}},
  note         = {{NUTS} 2021 version, 1:10M scale, {EPSG}:3035, GeoPackage format. Accessed: 2026-03-13}
}

@article{Monteiro2019Spatial,
title={Spatial Disaggregation of Historical Census Data Leveraging Multiple Sources of Ancillary Information},
author={J. Monteiro and B. Martins and Patricia Murrieta-Flores and J. Pires},
journal={ISPRS Int. J. Geo Inf.},
year={2019},
volume={8},
pages={327},
doi={10.3390/ijgi8080327}
}

@article{Sapena2022Empiric,
title={Empiric recommendations for population disaggregation under different data scenarios},
author={M. Sapena and Marlene Kühnl and M. Wurm and Jorge E. Patiño and J. Duque and H. Taubenböck},
journal={PLoS ONE},
year={2022},
volume={17},
doi={10.1371/journal.pone.0274504}
}

@article{Qiu2022Disaggregating,
title={Disaggregating population data for assessing progress of SDGs: methods and applications},
author={Y. Qiu and Xuesheng Zhao and Deqin Fan and Songnian Li and Yijing Zhao},
journal={International Journal of Digital Earth},
year={2022},
volume={15},
pages={2 - 29},
doi={10.1080/17538947.2021.2013553}
}

@article{Georgati2024Modeling,
title={Modeling population distribution: A visual and quantitative analysis of gradient boosting and deep learning models for multi‐output spatial disaggregation},
author={Marina Georgati and J. Monteiro and Bruno Martins and Carsten Keßler and Henning Sten Hansen},
journal={Transactions in GIS},
year={2024},
volume={28},
pages={130 - 153},
doi={10.1111/tgis.13130}
}

@article{Monteiro2018hybrid,
title={A hybrid approach for the spatial disaggregation of socio-economic indicators},
author={J. Monteiro and Bruno Martins and J. Pires},
journal={International Journal of Data Science and Analytics},
year={2018},
volume={5},
pages={189-211},
doi={10.1007/s41060-017-0080-z}
}

@ARTICLE{thurner2018pandapower,
    author={L. Thurner and A. Scheidler and F. Sch{\"a}fer and J. Menke and J. Dollichon and F. Meier and S. Meinecke and M. Braun},
    journal={IEEE Transactions on Power Systems},
    title={pandapower — An Open-Source Python Tool for Convenient Modeling, Analysis, and Optimization of Electric Power Systems},
    year={2018},
    month={Nov},
    volume={33},
    number={6},
    pages={6510-6521},
    doi={10.1109/TPWRS.2018.2829021},
    ISSN={0885-8950}}

\appendix
\section{Hyperparameter Robustness Check}
\label{app:sweep}

This appendix reports the numerical results behind the sensitivity analysis in Section~\ref{sec:sensitivity}.

\subsection{Post-Correction Intensity Sweep}
\label{app:sweep_post}

Post-correction is applied post-hoc to a pre-trained \gls{gnn} baseline; no retraining is required.
Each sweep point is evaluated across all $3 \times 4 = 12$ seed--fold combinations.
Tab.~\ref{tab:sweep_a} reports the $16$-region mean \gls{rmse} at each intensity level.

\begin{table}[htbp]
\centering
\caption{Post-correction intensity sweep on the \gls{gnn} base ($16$-region mean \gls{rmse}, $12$ seed--fold evaluations per point).
$\alpha$: \gls{ntl} scaling coefficient; $\gamma$: Proximity decay exponent; $\beta$: joint \gls{ntl}+Proximity scaling factor.
Intensity zero corresponds to the uncorrected \gls{gnn} baseline.}
\label{tab:sweep_a}
\small
\begin{tabular}{ccc}
\toprule
\textbf{\gls{ntl} ($\alpha$)} & \textbf{Proximity ($\gamma$)} & \textbf{Combined ($\beta$)} \\
\cmidrule(lr){1-1} \cmidrule(lr){2-2} \cmidrule(lr){3-3}
$\alpha$ \quad \gls{rmse} & $\gamma$ \quad \gls{rmse} & $\beta$ \quad \gls{rmse} \\
\midrule
0.00 \quad 9.27 & 0.0 \quad 9.27 & 0.00 \quad 9.27 \\
0.25 \quad 9.19 & 0.5 \quad 9.12 & 0.25 \quad 9.07 \\
0.50 \quad 9.20 & 1.0 \quad 8.79 & 0.50 \quad 8.97 \\
0.75 \quad 9.29 & 1.5 \quad 8.65 & 0.75 \quad 9.58 \\
1.00 \quad 9.45 & 2.0 \quad 9.42 & 1.00 \quad 11.20 \\
1.50 \quad 10.00 & 2.5 \quad 10.57 & 1.25 \quad 12.89 \\
2.00 \quad 10.81 & 3.0 \quad 11.49 & 1.50 \quad 14.22 \\
\bottomrule
\end{tabular}
\end{table}

For \gls{ntl}, the minimum \gls{rmse} ($9.19$) at $\alpha = 0.25$ represents a non-significant $0.9\%$ improvement over the baseline; \gls{rmse} increases monotonically beyond this point.
For Proximity, the minimum ($8.65$) at $\gamma = 1.5$ is a $6.7\%$ improvement, but this optimum does not transfer to the combined correction: the default $\gamma = 2.0$ already produces antagonism when \gls{ntl} is also applied.
For the combined sweep, any $\beta > 0.5$ produces \gls{rmse} worse than the baseline.

\subsection{Prior-Loss Weight Sweep}
\label{app:sweep_prior}

The prior-loss sweep requires retraining at each weight setting.
To limit computational cost, a single fold (seed~42, fold~1) covering four regions is used.
Tab.~\ref{tab:sweep_b} reports the mean \gls{rmse} at each $\lambda$.

\begin{table}[htbp]
\centering
\caption{Prior-loss weight sweep (seed~$42$, fold~1, 4 regions).
\gls{rmse} is the mean across the four test regions.}
\label{tab:sweep_b}
\small
\begin{tabular}{ccc}
\toprule
$\lambda$ & \textbf{\gls{ntl} prior \gls{rmse}} & \textbf{Proximity prior \gls{rmse}} \\
\midrule
0.01 & 10.48 & 10.30 \\
0.05 & 10.55 & 11.03 \\
0.10 & 10.50 & 10.98 \\
0.20 & 11.09 & 10.51 \\
0.50 & 10.78 & 10.69 \\
\bottomrule
\end{tabular}
\end{table}

The Proximity prior shows a wider gap between $\lambda = 0.01$ (10.30) and the default $\lambda = 0.05$ (11.03) than \gls{ntl} does (10.48 vs.\ 10.55).
This gap is dominated by a single fold that includes London (\gls{gnn} \gls{rmse}~$\approx 16.9$), the outlier identified in Section~5.6.
At $\lambda \geq 0.2$ the prior constraint overpowers the main loss and convergence degrades, confirming that the default sits in the functional operating range.

\end{document}